\documentclass[11pt]{amsart}

\usepackage{fullpage}

\usepackage{url}

\usepackage{amssymb}

\usepackage{cite}

\renewcommand{\a}{\alpha}

\renewcommand{\d}{\delta}
\newcommand{\D}{\Delta}
\newcommand{\e}{\varepsilon}
\newcommand{\f}{\varphi}
\newcommand{\s}{\sigma}

\renewcommand{\k}{\kappa}
\renewcommand{\l}{\lambda}

\renewcommand{\O}{\Omega}
\renewcommand{\o}{\omega}

\newcommand{\cC}{{\mathcal C}}
\newcommand{\cM}{{\mathcal M}}
\newcommand{\cT}{{\mathcal T}}

\newcommand{\cL}{{\mathcal L}}

\newcommand{\cU}{{\mathcal U}}
\newcommand{\cV}{{\mathcal V}}
\newcommand{\cP}{{\mathcal P}}

\newcommand{\cG}{{\mathcal G}}

\newcommand{\cD}{{\mathcal D}}

\newcommand{\cZ}{\mathcal Z}
 
\newcommand{\cY}{\mathcal Y}

\newcommand{\bR}{\mathbb R}

\newcommand{\bZ}{\mathbb Z}

\newcommand{\be}{\begin{equation}}
\newcommand{\ee}{\end{equation}}

\newcommand{\beaa}{\begin{eqnarray*}}
\newcommand{\bea}{\begin{eqnarray}}
\newcommand{\beal}[1]{\begin{eqnarray}\label{#1}}
\newcommand{\bean}{\begin{eqnarray}\nonumber}
\newcommand{\beadl}[1]{\begin{deqarr}\label{#1}}
\newcommand{\eeadl}[1]{\arrlabel{#1}\end{deqarr}}
\newcommand{\eeal}[1]{\label{#1}\end{eqnarray}}
\newcommand{\eead}[1]{\end{deqarr}}
\newcommand{\eea}{\end{eqnarray}}
\newcommand{\eeaa}{\end{eqnarray*}}

\renewcommand{\to}{\rightarrow}

\DeclareMathOperator{\Ima}{Im}

\DeclareMathOperator{\Ker}{Ker}

\renewcommand{\phi}{\varphi}
\renewcommand{\epsilon}{\varepsilon}
\renewcommand{\hat}{\widehat}

\newcommand{\<}{\langle}
\renewcommand{\>}{\rangle}

\newcommand{\w}{\widetilde}

\theoremstyle{plain}
\newtheorem{theorem}{Theorem}[section]

\newtheorem{remark}[theorem]{Remark}

\newtheorem{lemma}[theorem]{Lemma}

\newtheorem{proposition}[theorem]{Proposition}
\newtheorem{corollary}[theorem]{Corollary}

\theoremstyle{definition}

\def\endproof{\qed \medskip}
\def\blacksquare{\hbox to .60em {\vrule width .60em height .60em}}

\numberwithin{equation}{section}

\begin{document}

\title[ ]{On the conformal method for the Einstein constraint equations}

\author[ ]{Michael T. Anderson}

\address{Department of Mathematics, Stony Brook University, Stony Brook, N.Y.~11794-3651, USA} 
\email{anderson@math.sunysb.edu}
\urladdr{http://www.math.sunysb.edu/$\sim$anderson}

\thanks{Partially supported by NSF grant DMS 1607479} 

\begin{abstract}
In this work, we use the global analysis and degree-theoretic methods introduced by Smale to study the existence and 
multiplicity of solutions of the vacuum Einstein constraint equations given by the conformal method of 
Lichnerowicz-Choquet-Bruhat-York. In particular this approach gives a new proof of the existence result of 
Maxwell and Holst-Nagy-Tsogtgere. We also relate the method to the limit equation of Dahl-Gicquaud-Humbert 
and the non-existence result of Nguyen.

\end{abstract}

\maketitle

\setcounter{section}{0}
\setcounter{equation}{0}

\section{Introduction}

  Let $(M, g, K)$ be a triple consisting of a closed 3-manifold $M$, a Riemannian metric $g$ and a symmetric bilinear 
form $K$ on $M$. The constraint equations for the vacuum Einstein equations are given by 
\be \label{div0}
\d(K - Hg) = 0,
\ee
\be \label{scal}
|K|^{2} - H^{2} - R_{g} = 0,
\ee
where $\d$ is the divergence with respect to $g$, $H = tr_{g}K$ and $R_{g}$ is the scalar curvature of $(M, g)$. 
The equation \eqref{div0} is called the divergence or momentum constraint while \eqref{scal} is the Hamiltonian or 
scalar constraint. They are the Gauss-Codazzi and Gauss equations respectively of a hypersurface 
embedded in a $4$-dimensional Ricci-flat Lorentzian space-time $(\cM, g^{(4)})$. The space of solutions of 
the constraint equations \eqref{div0}-\eqref{scal} will be denoted by $\cC$. 

    The fundamental theorem of Choquet-Bruhat \cite{CB} guarantees that a smooth triple $(M, g, K)$ satisfying the constraints 
\eqref{div0}-\eqref{scal} form an initial data or Cauchy hypersurface of a space-time solution $(\cM, g^{(4)})$ of the 
vacuum Einstein equations $Ric_{g^{(4)}} = 0$. The metric and second fundamental form of $g^{(4)}$ induced on $M$ 
are given by $(g, K)$. 

    The equations \eqref{div0}-\eqref{scal} are highly underdetermined; there are $4$ equations for the $12$ unknown 
components of $(g, K)$. A basic issue of interest has been to determine whether there is a natural space of ``free" 
or ``unconstrained" data $\cD$, formally with $8$ degrees of freedom, which upon specifying an element in $\cD$, 
reduce the equations \eqref{div0}-\eqref{scal} to a determined set of equations. Ideally, one would then be able to 
uniquely solve these equations, giving then an effective parametrization of the dynamical gravitational degrees of freedom, 
i.e.~the space $\cC$, from the data in $\cD$. 

   A priori there are of course many possible choices for the free data space $\cD$. One would like $\cD$ to be as simple as 
possible topologically. On the other hand, very little seems to be known about the topology of the space $\cC$ of solutions of 
the constraint equations. 
   
   By far the best understood and most well-studied choice, especially for the case of closed manifolds considered here, is 
that given by the conformal method of Lichnerowicz-Choquet-Bruhat-York, cf.~for instance \cite{BI}, \cite{CBY}, \cite{I}, or 
one of its variants \cite{BI}, \cite{Mx2}. For the conformal method, $\cD$ has the following product structure. Let $\cG$ 
be the space of (pointwise) conformal equivalence classes $[g]$ of $C^{\infty}$ smooth metrics $g$ on $M$ and let $\cT$ 
be the fibration over $\cG$ with fiber over $[g]$ given by the space of equivalence classes $[g, \s]$ of $C^{\infty}$ smooth 
symmetric 2-tensors $\s$ which are transverse-traceless with respect to $g$. Thus $\d_{g}\s = tr_{g}\s = 0$, where 
$(g, \s)$ is any representative of $([g, \s])$; the equivalence relation is given by $(g, \s) \sim (\psi^{4}g, \psi^{-2}\s)$, 
cf.~\cite{Mx2}. Next, let $C^{\infty}(M)$ denote the space of smooth scalar functions $H$ on $M$. Then $\cD$ 
(i.e.~$\cD_{C^{\infty}}$) is given by 
$$\cD = \cT \times C^{\infty}(M).$$
It is easily verified that $\cD$ has formally $8$ degrees of freedom and is contractible. The space $\cD$ is commonly called the 
space of seed data for the conformal method. 

  Roughly speaking, given a pair $(g, K)$, $[g]$ represents the conformal class of $g$, $\s$ represents the 
transverse-traceless part of $K$ with respect to $g$, and $H$ represents the mean curvature $H = tr_{g}K$. The remaining 
degrees of freedom are then a conformal factor $\f$ for the metric and a vector field $X$ for the action of diffeomorphisms on 
symmetric bilinear forms. Given a choice of background metric $g_{0}$ for the conformal class $[g]$, (which then breaks the 
conformal symmetry), the constraint equations \eqref{div0}-\eqref{scal} give a determined system of equations for $(\f, X)$, 
cf.~\eqref{div}-\eqref{lich} below. The basic issue is then understanding the existence and uniqueness of such solutions. 
However, for the conformal method, the data $(\f, X)$ should be determined from the conformal data $([g, \s])$, 
i.e.~the structure of the set of solutions $(\f, X)$ should be independent of the choice of representative $g_{0} \in [g]$.  

  Let $\cC$ be the space of all $C^{\infty}$ smooth pairs $(g, K)$ on $M$ satisfying the constraint equations \eqref{div0}-\eqref{scal}. 
Instead of studying the solvability of the equations \eqref{div0}-\eqref{scal} for fixed data $([g, \s], H) \in \cD$, 
(given a choice of representative $g_{0} \in [g]$), we consider the behavior of the natural (projection) maps 
\be \label{Pi}
\begin{array}{cc}
\Pi_{\a}: \cC \to \cD, \\
\Pi_{\a}(g, K) = ([g, \s], H).
\end{array}
\ee
As first made clear by Maxwell \cite{Mx2}, such maps are only defined via an auxiliary choice of volume form $\a$ on $M$. 
The choice of $\a$ gives an identification of the tangent space $T_{[g]}\cG$ with the cotangent space $T_{[g]}^{*}\cG$ via the pairing 
$$\<k_{\a}(u), v\> = \int_{M}\<u, v\>_{g}\a,$$
for $g \in [g]$. This identification is necessary since the trace-free part $K_{0}$ of $K$ is a tangent vector, $K_{0} \in T_{[g]}\cG$, while 
$\s \in T_{[g]}^{*}\cG$ is in the cotangent space. The spaces $T\cG$ and $T^{*}\cG$ transform differently under conformal changes 
of the representative $g \in [g]$.  Thus in \eqref{Pi}, $[g]$ is the conformal class of $g$, $H = tr_{g}K$ and 
$$\s = P_{\a}(K_{0})$$
where $P_{\a}$ is the $L^{2}$ orthogonal projection with respect to $\a$ of $K_{0}$ onto 
the transverse-traceless tensors with respect to $g$. We refer to \cite{Mx2} for a detailed discussion of this issue.  

  We note that the maps $\Pi_{\a}$ depend continuously (and even smoothly in a natural sense) on $\a$. However, as 
discussed in \cite{Mx3}, there are examples where specific fibers of $\Pi_{\a}$, i.e.~the space of solutions $(\f, X)$ of 
the constraint equations, depend significantly on the choice of gauge $\a$. Recall also that two volume forms $\a$ and 
$\a'$ of the same total volume on $M$ are related by a diffeomorphism of $M$, i.e.~$\a' = \psi^{*}\a$, for some 
$\psi \in \mathrm{Diff}(M)$. 

  To describe the fibers of the map $\Pi_{\a}$, let $g_{0}$ be a representative metric in $[g]$ with volume form 
$dv_{g_{0}}$ and let 
\be \label{lapse}
N = \frac{1}{2}\frac{dv_{g_{0}}}{\a};
\ee
the scalar field $N$ is called the densitized lapse. Let $\s$ be a transverse-traceless (2,0) tensor with respect to $g_{0}$. 
One then forms $(g, K)$ by setting 
\be \label{g}
g = \f^{4}g_{0},
\ee
\be \label{K}
K = \f^{-2}(\s + \frac{1}{2N}\hat \cL_{X}g_{0}) + \frac{1}{3}H\f^{4}g_{0},
\ee
where $\hat \cL_{X}g_{0}$ is the conformal Killing operator with respect to $g_{0}$: $\hat \cL_{X}g_{0} = \cL_{X}g_{0} - 
\frac{2}{3}div_{g_{0}}X g_{0}$. The constraint equations \eqref{div0}-\eqref{scal} then become a coupled system of equations 
for $(\f, X)$ which take the form 
\be \label{div}
\d({\tfrac{1}{2N}}\hat \cL_{X}g_{0}) = -{\tfrac{2}{3}}\f^{6}dH,
\ee
for the divergence constraint while the Hamiltonian or scalar constraint takes the form of the Lichnerowicz equation 
\be \label{lich}
8\D \f = R_{0}\f - |\s + {\tfrac{1}{2N}}\hat \cL_{X}g_{0}|^{2}\f^{-7} + {\tfrac{2}{3}}H^{2}\f^{5}.
\ee
Here $\d$ and $\D$ are the divergence and Laplacian with respect to $g_{0} \in [g]$ and $R_{0}$ is the scalar curvature of 
$g_{0}$. 

  It is straightforward to verify (cf.~again \cite{Mx2}) that if $g_{1} = \psi^{4}g_{0}$, $\s_{1} = \psi^{-2}\s$ is a different representative of 
$([g,\s])$ and the gauge is changed so that $\a_{1} = \psi^{6}\a$, then solutions $(\f, X)$ with respect to $(g_{0}, \s, H, \a)$ 
transform exactly to solutions $(\psi^{-1}\f , X)$ with respect to $(g_{1}, \s_{1}, H, \a_{1})$. This shows the coupling between 
the choice of gauge $\a$ and choice of representative $g_{0} \in [g]$. 

   It is well-known that the equations \eqref{div}-\eqref{lich} form a determined elliptic system for $(\f, X)$, given $(g_{0}, \s, H)$, 
cf.~also \S 3. In much of the literature on the conformal method, when $g_{0}$ is fixed, the gauge $\a$ is chosen so that 
$N = \frac{1}{2}$, i.e. $\a = dV_{g_{0}}$. 

\medskip 

  The basic question is then, given a choice of gauge $\a$, for what free data $([g, \s], H) \in \cD$ are these equations 
solvable, or even better, uniquely solvable. A complete answer regarding existence and uniqueness is known in the CMC case 
where $H = const$, cf.~\cite{I}, based on work of \cite{L}, \cite{CBY}, \cite{OY} and others. The near-CMC case, where the 
derivative $dH$ is sufficiently small compared with $H$ is also almost fully understood, cf.~\cite{I2} and references therein 
for a recent survey. The far-from-CMC case has been shown to be much more difficult and much less is understood. Two of 
the major results in this regime are the result of Holst-Nagy-Tsotgerel \cite{HNT} and Maxwell\cite{Mx1}, and that of 
Dahl-Gicquaud-Humbert \cite{DGH}; these results are also discussed further below. The first fundamental non-existence 
result was proved by Nguyen \cite{Ngu}. An excellent view of the current state of understanding is given in \cite{DHKM}, 
which provides strong numerical evidence for a great deal of complexity in the space of solutions. 

  The reason for the simplification in the CMC case is well-known; in this case one may set $X = 0$ in \eqref{div} 
and the system \eqref{div}-\eqref{lich} reduces to the Lichnerowicz equation \eqref{lich} for $\f$ involving only the given 
data $R_{0}, \s, H$. The map $\Pi_{\a} = \Pi$ in \eqref{Pi} is then independent of $\a$. The Lichnerowicz equation is closely related 
to the well-understood Yamabe equation for constant scalar curvature metrics. 

\medskip 

  In this paper, we take a somewhat different perspective from previous work on this issue, namely a global analysis perspective 
going back to the work of Smale \cite{Sm}. As will be seen in \S 3, although the spaces $\cC$ and $\cD$ are not smooth 
manifolds globally, they do have formal tangent spaces $T_{(g,K)}\cC$, $T_{([g,\s],H)}\cD$ everywhere. The linearization 
$$D\Pi_{\a}: T_{(g,K)}\cC \to T_{\Pi(g,K)}\cD,$$ 
is a Fredholm map, of Fredholm index zero. 
  
  The main interest is the global behavior of the maps $\Pi_{\a}$. In particular, one would like to understand the image of $\Pi_{\a}$ 
and the injectivity of $\Pi_{\a}$, corresponding to the existence and uniqueness of solutions of \eqref{div}-\eqref{lich}. 
On the CMC class where $H = const$, i.e. 
$$\cC^{cmc} = \cC \cap \{H = const\},$$
as noted above, the restricted map 
$$\Pi^{cmc} = \Pi |_{\cC^{cmc}}: \cC^{cmc} \to \cD^{cmc} = \cD\cap \{H = const\}$$ 
is independent of $\a$. Although $\Pi^{cmc}$ does not surject onto $\cD^{cmc}$, its image $\Ima \Pi^{cmc}  \subset \cD^{cmc}$ 
and its injectivity are fully understood, cf.~also \S 2. 

 Returning to the general situation regarding \eqref{Pi}, the key to understanding global properties of $\Pi$ is to understand 
in what regions (if any) $\Pi$ is proper. Recall that a continuous map $F: X \to Y$ between topological spaces is proper if 
$F^{-1}(K)$ is compact in $X$, for any compact set $K \subset Y$. This issue is essentially equivalent to the existence of 
apriori estimates for solutions of the constraint equations \eqref{div}-\eqref{lich}. 
 
 Let $\cG' \subset \cG$ be the space of conformal classes which have no (non-zero) conformal Killing field, so that 
$$\cG' = \{[g] \in \cG: {\rm Ker} \hat \cL g_{0} = 0\},$$
for any $g_{0} \in [g]$. Let 
\be \label{d'}
\cD' \subset \cD
\ee
be the restriction of the fibration $\cD$ to the domain $\cG'$ and let 
$$\cC' = \Pi_{\a}^{-1}(\cD') \subset \cC,$$
with the induced map 
\be \label{Pi'}
\Pi_{\a}' : \cC' \to \cD'.
\ee  
We note that for trivial reasons,  $\Pi_{\a}$ in \eqref{Pi} is not expected to be proper over the region $\cD \setminus \cD'$. 
Namely, if $Z$ is a conformal Killing field in the conformal class $[g]$ and $(\f, X)$ solve the constraint equations 
\eqref{div}-\eqref{lich}, then so do $(\f, X + Z)$. This shows that one has control on $X$ only modulo the space of 
conformal Killing fields. Although only the term $\hat \cL_{X}g_{0}$ enters the constraint equations, the presence of 
low eigenvalues for the operator $\d (\frac{1}{2N}\hat \cL_{X}g_{0})$ suggests $\cL_{X}g_{0}$ cannot be controlled in general near 
a class $[g_{0}]$ with conformal Killing fields. For this reason, we essentially restrict to the map \eqref{Pi'} 
throughout this paper. It is well-known that the presence of conformal Killing fields 
causes difficulties in the conformal method, cf.~in particular the discussion in \cite{HMM}. These difficulties bear 
some relation with the classical Nirenberg problem of prescribed Gauss or scalar curvature, for metrics conformal 
to the standard round sphere $S^{n}(1)$. 

\medskip 

   Now Smale \cite{Sm} proved that proper Fredholm maps $F: X \to Y$ of index zero between separable Banach manifolds have a 
well-defined (mod 2) degree, $deg_{{\bf Z}_{2}}F$, the Smale degree, given by the cardinality (mod 2) of the fiber $F^{-1}(y)$, 
for any regular value $y \in Y$ of $F$. If $deg_{{\bf Z}_{2}}F = 1$, then $F$ is surjective. The approach in this work is to study 
the application of these ideas to the maps $\Pi_{\a}'$ in \eqref{Pi'}. 

  However, it is very difficult to understand in what regions $\Pi_{\a}'$ is proper or the cause of non-proper or divergent behavior 
in $\cC'$ with respect to $\Pi_{\a}'$. As an aid in this issue, it will be useful to choose a family of hypersurfaces of $\cC'$ on which 
$\Pi_{\a}'$ is more well-controlled. There are a number of possible choices, but for convenience we choose the following: for a 
given $p > 1$ consider the functional 
$$F_{p}: \cC' \to \bR^{+},$$
$$F_{p}(\f) = \int_{M}\f^{p}dv_{g_{0}}.$$
This requires a choice of background representative $g_{0} \in [g]$. Throughout the paper, we choose $g_{0}$ to be 
the unit volume Yamabe representative for $[g]$. Such metrics are unique, and vary smoothly with the conformal 
class $[g]$, for an open-dense set $\cU_{0}$ of conformal classes, cf.~\cite{An}. It follows that $F_{p}$ is a well-defined continuous 
function on the open-dense set over $\cU_{0}$ in $\cC'$ and bounded on the complement $\cC'\setminus \cU_{0}$. Hence, it may 
be mollified in a neighborhood of $\partial \cU_{0}$ to give a continuous function on $\cC'$, cf.~\S 3. 

  There is not a unique choice for $p$ but for convenience, we choose $p = 8$ and consider the level sets of $F_{8}$. Thus let 
$$\cC^{\o} = \{(g, K) \in \cC': \int_{M}\f^{8}dv_{g_{0}} = \o\} \subset \cC'.$$ 
The maps $\Pi_{\a}'$ in \eqref{Pi'} restrict to give maps 
$$(\Pi_{\a}')^{\o}: \cC^{\o} \to \cD'.$$
All results of this paper hold for any choice of $\a$. In the following, to simplify the notation we denote by $\Pi$ any of the maps 
$\Pi_{\a}'$ above. 

  The first main result of this paper is the following: 
\begin{theorem}
For each $\o \in (0, \infty)$, (and each $\a$), the map 
\be \label{Pi+}
\Pi^{\o} = \Pi|_{\cC^{\o}}: \cC^{\o} \to \cD'
\ee
is a continuous, proper map. 
\end{theorem} 
Theorem 1.1 implies in particular that for given data $([g, \s], H) \in \cD'$, the set of solutions 
$(\f, X)$ to the constraint equations \eqref{div}-\eqref{lich} in $\cC^{\o}$ is compact. 

 Formally the map $\Pi^{\o}$ is a Fredholm map, of Fredholm index $-1$, and so it is natural to study the intersection 
properties of the image $\Ima \Pi^{\o}$ with 1-dimensional submanifolds, i.e.~curves, in $\cD'$. It follows from Theorem 
1.1 that for any properly embedded curve $L : \bR \to \cD'$, and any compact interval $I \subset \bR$, the intersection 
$$\Ima \Pi^{\o} \cap L(I),$$
is compact; equivalently the inverse image $\cC^{\o}\cap \Pi^{-1}(L(I))$ is compact in $\cC^{\o}$. Generically (when $L$ is 
transverse to $\Pi^{\o}$), the intersection is a finite number of points: 
$$\# (\Ima \Pi^{\o} \cap \Ima (L(I)) < \infty.$$
To obtain a well-defined intersection number, one needs to strengthen the statement above to the statement that the full 
intersection 
$$\Ima \Pi^{\o} \cap \Ima L,$$
is compact. 

   There are a number of natural choices for such curves $L$. In this paper, we restrict to only one choice closely related 
to previous studies of the conformal method. Thus consider lines in the space of transverse-traceless tensors 
$\s$, i.e.~lines of the form 
$$L_{\s}(\l) = ([g, \l \s], H) \in \cD',$$
with $([g, \s], H)$ fixed. 

\begin{theorem}
The intersection 
\be \label{int1}
\Ima \Pi^{\o} \cap \Ima L_{\s},
\ee
is compact, for any $([g, \s], H) \in \cD'$ and generically the intersection \eqref{int1} consists of a finite number of points. 
There is a well-defined $\bZ_{2}$-intersection number 
$$I_{\bZ_{2}}(\o, \{g, \s\}) = \# \{(\Pi_{\e}^{\o})^{-1}(L_{\s})\}, \ \ (mod \, 2),$$ 
independent of $\o$, $\e$, and the data $([g, \s], H)$, and 
$$I_{\bZ_{2}}(\o, \{g, \s\}) = 0.$$
\end{theorem} 

   The map $\Pi_{\e}^{\o}$ is an $\e$-perturbation or regularization of $\Pi^{\o}$, cf.~\eqref{Ce}, \eqref{Piwe2} for the 
exact definition. A more precise version of Theorem 1.2 is given in Theorem 6.1. The intersection number $I_{\bZ_{2}}(\o, \{g,\s\})$ 
corresponds roughly to the Smale $\bZ_{2}$-degree of the map 
\be \label{p}
\w \Pi^{\o}: \cC^{\o} \to \cP',
\ee
where $\cP'$ is the projectivization of $\cD'$, i.e.~$\cP' = (\cD' \setminus Z)/\sim$ where the equivalence classes $\{g, \s\}$ are 
given by $([g, \s], H) \sim ([g, \l \s], H)$ and $Z$ is the zero section $([g, 0], H)$. However, the map \eqref{p} is not quite proper 
due to the singular behavior near the zero-section $Z$. 

  The transversality and intersection number properties discussed above require smooth separable Banach manifold structures on 
the domain $\cC'$ and target $\cD'$ spaces. In the case of the constraint space $\cC$, this is the issue of linearization 
stability of solutions of the Einstein equations, studied in detail by Fischer, Marsden and Moncrief, cf.~\cite{FM2}, 
\cite{FMM}, \cite{Mo}. One has the decomposition 
\be \label{rs}
\cC = \cC^{reg} \cup \cC^{sing}, 
\ee
corresponding to the regions where $0$ is a regular or singular value of the constraint map. This gives a $C^{\infty}$ smooth 
(Frechet) manifold structure to the regular region $\cC^{reg}$. The space $\cC^{sing}$ consists of Killing initial data $(g, K)$, 
(for which the vacuum development has a non-zero Killing field). There is a basic conjecture, cf.~\cite{BCS}, that $\cC^{reg}$ 
is open and dense in $\cC$. While this is known to be true in the CMC case, cf.~\cite{FM1}, \cite{BCS}, this remains an open 
problem in general. 
 
   However, as pointed out by Bartnik in \cite{B}, the proof in \cite{FM2}, \cite{FMM}, \cite{Mo} cannot be adapted to give a finite 
differentiablity or Banach manifold structure to $\cC^{reg}$. Based on the conformal method, we prove in \S 3, cf.~Theorem 3.1, 
that $\cC^{reg}$ can be given a separable Banach manifold structure. The singular set $\cC^{sing}$ will be regularized to a 
smooth Banach manifold structure by considering the space of solutions to the $\e$-perturbed constraint equations $\cC_{\e}$; 
this is carried out in detail in \S 3. 

\medskip 

  Theorem 1.2 shows that solutions $(\f, X)$ of the constraint equations in any given level $\cC^{\o}$ over the line $\{\s\} = \{\l \s\}$ 
(with $([g, \s], H)$ fixed) typically come in pairs, or there are no solutions over $\{\s\}$. 
One sees this very easily in the CMC case, where $X = 0$ and $\l$ only appears as $\l^{2}$ in the Lichnerowicz equation 
\eqref{lich}. Thus when $H = const$, $(\f, 0) \in \cC^{\o}$ is a solution with data $([g, \l \s], H)$ if and only if it is also a solution with data 
$([g, -\l \s], H)$. 
 
  Next we study the behavior of solutions $(\f, X) \in \cC^{\o}$ as $\o$ varies over $\bR^{+}$. 
Consider the map 
$$\w \Pi: \cC' \to \cP',$$
as in \eqref{p}, without the restriction to $\cC^{\o}$. 
In the smooth or regular region, $\w \Pi$ is a smooth Fredholm map of index one. Thus, choose (for instance) a regular 
value $(\{g, \s\}, H)$ of $\w \Pi$. Let $L_{\s} = \{([g, \l \s], H): \l \in \bR\}$ be the line forming the equivalence class of 
$(\{g, \s\}, H)$. 
Then the inverse image 
$$\Gamma = \w \Pi^{-1}(\{g, \s\}, H)$$ 
is a collection of curves ($1$-manifolds)  $\{\ell(t) = (\f(t), X(t)\}$ mapping to the line $L_{\s}$. The intersection of 
$\Gamma$ with any level set $\cC^{\o}$ of $F$ is compact, and generically an even number (possibly zero) of points. 
It follows that $\Gamma$ is a collection of embedded circles $S^{1}$ or properly embedded arcs $\sim \bR$ in $\cC'$. 
As will be seen below, a special role is played by the value $\l = 0$ on $\Gamma$.  Let 
$$\cC_{\o_{0}}^{\o_{1}} = \cC' \cap \{\f: \o_{0} \leq F(\f) \leq \o_{1}\},$$
and let $Y[g]$ denote the Yamabe constant of the conformal class $[g]$. 

\begin{theorem}
Suppose $Y[g] > 0$. Given any line $([g, \l \s], H)$, $\s \neq 0$ (with $([g, \s], H)$ arbitrary), there is an $\o_{0}$, depending on 
$([g, \s], H)$, such that 
$$\Gamma_{\o_{0}} := \Gamma \cap \cC_{0}^{\o_{0}},$$ 
is a pair of disjoint arcs $(\f_{\pm}(t), X_{\pm}(t))$, $t \in (0,t_{0}]$. The level parameter $\o \in (0, \o_{0}]$ is a smooth parametrization 
of $\Gamma_{\o_{0}}$. One has $\l > 0$ on $\Gamma_{+}$, $\l < 0$ on $\Gamma_{-}$ with $|\l|$ monotone increasing with 
$\o$ on $\Gamma_{\pm}$ and 
$$\Pi(\Gamma_{\pm}) = [\l_{-}, 0) \cup (0, \l_{+}].$$
There is no solution in $\cC_{0}^{\o_{0}}$ with $\l = 0$. 

 If $Y[g] < 0$, then 
$$\Gamma \cap \cC_{0}^{\o_{0}} = \emptyset,$$
i.e.~there are no solutions of the constraint equations with $\o$ sufficiently small. 
\end{theorem} 

  Again we refer to Theorem 6.3 and Corollary 6.5 for a more precise statement of Theorem 1.3. 

  Theorem 1.3 gives the existence of solutions $(\f, X)$ of the constraint equations with data $([g, \l \s], H)$ 
for $Y[g] > 0$, $\l$ sufficiently small, $\s \neq 0$ and $([g, \s], H)$ arbitrary; moreover, such solutions have $\o$ small, 
and so also have small volume and are the unique solutions with $\o$ small. This existence result was previously proved by 
Maxwell \cite{Mx1}, cf.~also Holst-Nagy-Tsotgerel \cite{HNT}. The proof of Theorem 1.3 is quite different than these approaches. 

  The transition between the existence and non-existence of solutions with small $\o$ in passing from $Y[g] > 0$ through 
$Y[g] = 0$ to $Y[g] < 0$ is quite subtle, cf.~Remark 6.6. 

  We conclude the paper with a discussion of the large-scale, i.e.~large $\o$, behavior of solutions in $\cC'$; this is governed by 
the ``limit equation" of Dahl-Gicquaud-Humbert \cite{DGH}: 
\be \label{limit0}
\d ({\tfrac{1}{2N}}\hat \cL_{\bar X}g_{0}) = -\sqrt{{\tfrac{2}{3}}}|{\tfrac{1}{2N}}\hat \cL_{\bar X}(g_{0})|\frac{dH}{H}.
\ee
This is discussed further in Proposition 6.7 and together with the results above leads to the following: 

\begin{theorem}
Let $\O$ be a domain in $\cG'\times C^{\infty}(M)$ with $Y[g] > 0$ and $H > 0$ and suppose the 
limit equation \eqref{limit} has no non-zero solution for $([g], H) \in \O$. 

  Then for any $\s \neq 0$ there is a solution $(\f, X)$ of the constraint equations \eqref{div}-\eqref{lich} over the data 
$([g, \s], H)$ with $([g], H) \in \O$. 
\end{theorem}

  We refer to Corollary 6.8 for a more detailed statement of this result. 

\medskip 

  The contents of the paper are briefly as follows. In \S 2, we introduce background material and results needed for 
the work to follow. We also summarize the known existence and uniqueness results for CMC solutions and prove that the map 
$\Pi^{cmc}$ is a diffeomorphism in a neighborhood of $\cC^{cmc}$, cf.~Theorem 2.1. This leads to a simple proof of previous 
near-CMC results in many cases, cf.~Corollary 2.2.  In \S 3, we study the constraint map 
and prove the Banach manifold structure results for the vacuum and $\e$-perturbed vacuum solutions to the constraint equations, 
cf.~Theorem 3.1. It is also proved that the target space $\cD$ is a Banach manifold away from data admitting conformal Killing fields. 
The basic initial a priori estimates for the map $\Pi$ are derived in \S 4. Theorem 1.1 is proved in \S 5 together with some 
initial estimates on the behavior of solutions $(\f, X)$ with small $\o$. Theorems 1.2, 1.3 and 1.4 are then proved in \S 6. 

\medskip 

  I am very grateful to David Maxwell for pointing out an error in a previous version of this paper and for his invaluable help in 
explaining the current state-of-the-art of the conformal method. My thanks also to The-Cang Nguyen for his interest and 
correspondence.

\section{Background and Preliminary Material} 

  In this section, we present background material needed for the work to follow. Throughout the paper, $M$ denotes a 
compact $3$-manifold, without boundary. (All of the results of this work hold with only minor changes in higher 
dimensions $dim \, M \geq 3$). 
  
  To begin, we discuss the topology of the spaces $\cC$ and $\cD$. The $C^{\infty}$ topology is a Fr\'echet space 
topology, which is not suitable for analysing nonlinear Fredholm maps, mainly due to the failure of the inverse function 
theorem. 

   The simplest Banach spaces on which elliptic operators are well-behaved are the H\"older spaces $C^{m,\a}$ and 
Sobolev spaces $W^{k,p}$ for suitable $(m, \a)$ or $(k, p)$. For the Einstein evolution equations where energy estimates 
play a key role, one usually uses the Sobolev spaces $H^{s} = W^{s,2}$, for suitable $s \geq 2$. However, we will use the 
H\"older spaces $C^{m,\a}$ here, since the projection map $\Pi$ in \eqref{Pi} is only known to be well-behaved in H\"older 
spaces $C^{m,\a}$; this is discussed further in \S 3. (It is possible one could work in the class of Morrey spaces \cite{Ad}, 
but this will not be pursued here). Throughout the paper we assume $m \geq 2$, $\a \in (0,1)$.  (We will not be concerned 
with obtaining the lowest possible regularity results). 

  Moreover, it is well-known that H\"older spaces $C^{m,\a}$ are not separable Banach spaces; they do not admit a 
countable basis. Since separability will be an important property, we work instead with a maximal closed separable 
subspace of $C^{m, \a}$, namely the so-called little H\"older space $c^{m,\a}$. This may be defined to be the 
completion of $C^{m+1}$ or $C^{\infty}$ with respect to the $C^{m,\a}$ norm. Equivalently, functions $f$ on smooth 
domains $\Omega \subset \bR^{n}$ are in $c^{0,\a}(\Omega)$ if $f \in C^{0,\a}(\Omega)$ and, for $x, y \in \Omega$, 
$$\lim_{r\to 0}\sup_{0 < dist(x,y)<r}\frac{|f(x) - f(y)|}{dist(x,y)^{\a}} = 0.$$
The space $c^{m,\a}(\Omega)$ consists of functions $f$ whose partial derivatives up to order $m$ exist and are in 
$c^{0,\a}(\Omega)$. The space $c^{m,\a}$ is a separable Banach space, embedded as a closed subspace of $C^{m,\a}$, 
cf.~\cite{BL}. Note that $C^{m,\a'} \subset c^{m,\a}$ for all $\a' > \a$.

\medskip 

   Let $Met^{m,\a}(M)$ be the space of $c^{m,\a}$ metrics $g$ on $M$; thus in a smooth atlas for $M$, the coefficients 
of $g$ are $c^{m,\a}$ functions. Similarly let $S_{2}^{m-1,\a}(M)$ be the space of $c^{m-1,\a}$ symmetric bilinear 
forms $K$ on $M$. Define then 
$$\cC : = \cC^{m,\a} \subset Met^{m,\a}(M)\times S_{2}^{m-1,\a}(M)$$
to be the subspace satisfying the constraint equations \eqref{div0}-\eqref{scal}, with the induced topology. 

  Next, let $\cG^{m,\a}$ be the space of $c^{m,\a}$ conformal equivalence classes of metrics in $Met^{m,\a}(M)$; thus 
$g_{1} \sim g_{2}$ if $g_{2} = \f^{4}g_{1}$, for some positive function $\f \in c^{m,\a}$. Let $\cT^{m-1,\a}$ be the fibration 
of $c^{m-1,\a}$ transverse-traceless tensors $\s$ over $\cG^{m,\a}$; thus $\cT^{m-1,\a}$ consists of pairs $(g, \s)$ with 
$tr_{g}\s = \d_{g}\s = 0$, modulo the equivalence relation $(g, \s) \sim (\psi^{4}g, \psi^{-2}\s)$, cf.~\cite{Mx2} 
for details. Define also 
$$\cD := \cD^{m-1,\a} = \cT^{m-1,\a}\times c^{m-1,\a}(M).$$

  Thus, for each choice of $\a$, we have the map $\Pi$ as in \eqref{Pi}, 
\be \label{Pi2}
\Pi: \cC \to \cD.
\ee 
As discussed in the Introduction, the subscript $\a$ is dropped for notational simplicity. 

   The spaces $\cC$ and $\cD$ are not globally smooth manifolds and it will be 
important to understand the structure of the domains $\cC^{reg}$ and $\cD^{reg}$ in $\cC$, $\cD$ which are 
smooth manifolds. The space $\cG$ is not a manifold at the points $[g]$ which admit a conformal Killing field. 
For both this reason and the fact that $\Pi$ is not proper over such conformal classes, as in \S 1 we will generally 
restrict to the map $\Pi'$ as in \eqref{Pi'}. In \S 3 we discuss the manifold regions of $\cC$ and $\cD$; the singular 
region $\cC^{sing}$ of $\cC$ will be analysed by using a simple perturbation or regularization to ``near" solutions 
of the vacuum constraint equations.

\medskip 

   Next we discuss the results established in the CMC case where $H = const$, cf.~\cite{I}, \cite{Mx2}. Let 
$\cD^{cmc} \subset \cD'$ be the subset of $([g, \s], H)$ where $H = const$. Although the main existence and 
uniqueness results discussed below hold for $\cD^{cmc} \subset \cD$, we exclude the data $[g]$ which contain 
conformal Killing fields; it then follows from Proposition 3.3 below that $\cD^{cmc}$ is a smooth Banach manifold. 
Let $Y[g]$ be the Yamabe constant of $[g]$ and let 
\be \label{dcmc}
\begin{array}{lll}
\cD_{+}^{cmc} = \{([g, \s], H) \in \cD^{cmc}: Y[g] > 0, \s \neq 0\},\\
\cD_{0}^{cmc} = \{([g, \s], H) \in \cD^{cmc}: Y[g] = 0,  \s \neq 0 \ {\rm and} \ H \neq 0\},\\ 
\cD_{-}^{cmc} = \{([g, \s], H) \in \cD^{cmc}: Y[g] < 0, H \neq 0\}, 
\end{array}
\ee
where $\s$ or $H \neq 0$ denotes $\s$  or $H$ is not identically zero. Setting 
$$\cD_{ex}^{cmc} = \cD_{+}^{cmc} \cup \cD_{0}^{cmc} \cup \cD_{-}^{cmc},$$
($ex$ is meant to denote `exists'), one sees that $\cD_{ex}^{cmc}$ is a connected, open subset of $\cD^{cmc}$. 

  Let $\cC_{ex}^{cmc} = (\Pi')^{-1}(\cD_{ex}^{cmc})$. Then the map 
\be \label{Piex}
\Pi_{ex}^{cmc}: \cC_{ex}^{cmc} \to \cD_{ex}^{cmc},
\ee
is a smooth, proper homeomorphism; in particular $\Pi_{ex}^{cmc}$ is one-to-one and onto. (We recall that 
$\Pi^{cmc}$ is independent of $\a$).  

  Let $\cD_{nx}^{cmc} = \cD^{cmc}\setminus \cD_{ex}^{cmc}$ be the complementary closed set, ($nx$ is meant to 
denote `non-existence'), so that 
$\cD_{nx}^{cmc}\subset \cD^{cmc}$ is given by 
\be \label{cmc-}
\cD_{nx}^{cmc} = \left\{ \begin{array}{lll} 
Y[g] < 0: & H = 0, \\
Y[g] = 0: &  \s = 0   \ {\rm or} \  H = 0, \\
Y[g] > 0: & \s = 0.
\end{array}
\right.
\ee
Correspondingly, let $\cC_{nx}^{cmc} = \Pi^{-1}(\cD_{nx}^{cmc})$. Then 
$$\Pi_{nx}^{cmc}: \cC_{nx}^{cmc} \to \cD_{nx}^{cmc},$$
is the empty map, i.e.~$\cC_{nx}^{cmc} = \emptyset$, except in the exceptional, boundary, situation where $Y[g] = 0$, 
$\s = H = 0$ in which case one has the trivial solutions $(\f, X) = (const, 0)$ with $g = c^{4}g_{0}$ scalar-flat 
metrics with $K = 0$. 

  This gives a very clear distinction between the regions of existence and non-existence of solutions of the constraint 
equations \eqref{div}-\eqref{lich}. The map $\Pi_{ex}^{cmc}$ must thus degenerate essentially everywhere on approach to  
$\partial \cC_{ex}^{cmc}$, where the boundary is taken as a subset of $Met^{m,\a}(M)\times S_{2}^{m-1,\a}(M)$. Since 
$X = 0$, this means that $\f$ must degenerate, as a positive function in $c^{m,\a}$, on approach to essentially any 
point in $\partial \cC_{ex}^{cmc}$. This will be seen in further detail in the analysis in \S 4. 

\medskip 

  It follows from Theorem 3.1 below, (cf.~\eqref{cmcreg}), that the space $\cC_{ex}^{cmc}$ is a smooth Banach manifold, 
so that $\Pi_{ex}^{cmc}$ in \eqref{Piex} is a smooth map of Banach manifolds. 

\begin{theorem}
The map $\Pi_{ex}^{cmc}$ in \eqref{Piex} is a smooth diffeomorphism. 
\end{theorem} 

{\bf Proof:} Since $\Pi_{ex}^{cmc}$ is a smooth homeomorphism between Banach manifolds, it suffices to prove that $\Pi$ has no 
critical points at solutions $(\f, X)$ with data $([g, \s], H) \in \cD_{ex}^{cmc}$. Thus, suppose $(\f', X')$ is a solution of the linearized 
constraint equations \eqref{div}-\eqref{lich} with fixed data $([g, \s], H)$, so that $g_{0}' = \s' = H' = 0$. Here $g_{0}$ is chosen 
to be a unit volume Yamabe metric realizing $Y[g_{0}]$, cf.~also \S 3. Since $H = const.$, 
the linearized divergence constraint gives 
$$\d ({\tfrac{1}{2N}}\hat \cL_{X'}g_{0}) = 0.$$
Pairing this with $X'$ and applying the divergence theorem, it follows easily that $X' = 0$. The linearized Lichnerowicz 
equation then gives 
\be \label{linear}
8\D \f' = R_{0}\f' + 7|\s|^{2}\f^{-8}\f' + {\tfrac{10}{3}}H^{2}\f^{4}\f'.
\ee
If $R_{0} \geq 0$, all coefficients of $\f'$ on the right in \eqref{linear} are positive and it follows from the maximum principle 
that $\f' = 0$. This completes the proof in case $Y[g] \geq 0$. 

  Next suppose $Y[g] < 0$. Evaluating the Lichnerowicz equation \eqref{lich} at a point $p$ realizing $\min \f$ gives 
$R_{0}\f + {\tfrac{2}{3}}H^{2}\f^{5} \geq 0$ at $p$, so that 
$$R_{0} + {\tfrac{2}{3}}H^{2}\f^{4} \geq 0$$
for all $x \in M$, since $H$ and $R_{0}$ are constant. Substituting this in \eqref{linear} gives 
$$8\f'\D \f' \geq 7|\s|^{2}\f^{-8}(\f')^{2} + {\tfrac{8}{3}}H^{2}\f^{4}(\f')^{2}.$$
Since the right side here is non-negative, it follows for instance by integration by parts that $\f' = 0$. This completes the 
proof when $Y[g] < 0$. 

{\endproof}

  The next result gives a simple proof of the existence of solutions of the constraint equations near CMC data, i.e.~the 
CMC existence results above are stable under perturbation in $H$. 

\begin{corollary}
The map $\Pi$ is a diffeomorphism in a neighborhood of $\cC^{cmc} \subset \cC'$. In particular, for any data $([g], \s, H)$ with 
$H$ close to a constant, and satisfying the conditions \eqref{dcmc}, there is a unique solution $(\f, X)$ to the constraint 
equations near the corresponding CMC solution. 

\end{corollary}

{\bf Proof:} This is an immediate consequence of the proof of Theorem 2.1 and the inverse function theorem. 

{\endproof}

  For the work to follow in later sections, we recall here some basic facts from global analysis on separable Banach manifolds first 
developed by Smale \cite{Sm}. Let $F: X \to Y$ be a smooth map between connected separable Banach manifolds $X$, $Y$. The map 
$F$ is Fredholm if for each $x \in X$, the linearization $D_{x}F: T_{x}X \to T_{F(x)}Y$ is a Fredholm map, i.e.~$D_{x}F$ has finite 
dimensional kernel and cokernel, with $D_{x}F$ of closed range. A point $x \in X$ is a regular point of $F$ if the linearization $D_{x}F$ 
is a surjective bounded linear map. A point is a singular point if it is not a regular point. A point $y \in Y$ is a regular value of $F$ if 
every point in the inverse image $F^{-1}(y)$ is a regular point; otherwise $y$ is a singular value. By the Sard-Smale theorem 
\cite{Sm}, the regular values of $F$ are of second category in $Y$, so given as the intersection of a countable collection of 
open and dense sets in $Y$. Note that by definition, any point $y \notin \Ima F$ is a regular point of $F$. 

  If the Fredholm index of $F$ is zero, $F$ is a local diffeomorphism in a neighborhood of any regular point. If $y \in Y$ 
is a regular value of $F$, the inverse image $F^{-1}(y)$ is a discrete, countable collection of points in $X$.  
 
  Next, let $V$ be a compact connected finite dimensional manifold, possibly with boundary, of dimension at least one. Then for 
any $\e > 0$, any smooth embedding $g: V \to Y$ admits a smooth perturbation $g': V \to Y$, $\e$-close to $g$, such 
that $g'$ is transverse to $F$; this means that for any $(x, v) \in X\times V$ such that $F(x) = g'(v) = y$, $T_{y}Y$ is spanned 
by the image of $D_{x}F$ and $D_{v}g'$; 
$$T_{y}Y = \Ima D_{x}F + \Ima D_{v}g'.$$
In addition, for such maps $g'$ transverse to $F$, the inverse image $F^{-1}(g'(V))$ is a smooth embedded submanifold of $X$ 
of dimension equal to $dim\, V + index \, F$. 

  The results above do not require that $F$ is a proper Fredholm map. If $F$ is proper, then the regular values of $F$ are open 
and dense in $Y$. For any $y \in Y$, the inverse image $F^{-1}(y)$ is compact and for $y$ a regular value, the inverse image 
$F^{-1}(y)$ is a finite collection of connected manifolds of dimension $index \, F$. 

  Any proper Fredholm map $F: X \to Y$ of index zero has a well-defined (mod 2) degree, the Smale degree 
$$deg_{\bZ_{2}}F \in \bZ_{2},$$
defined as the cardinality (mod 2) of the inverse image $F^{-1}(y)$ for $y$ any regular value of $F$. We recall briefly the 
proof that $deg_{\bZ_{2}}F$ is well-defined. If $y, y'$ are two regular values of $F$, consider the inverse images $F^{-1}(y)$, 
$F^{-1}(y')$, each a finite set of points. Let $y(s)$ be a smooth path in $Y$, 
transverse to $F$, with endpoints $y, y'$. The inverse image $F^{-1}(y(s))$ is a finite collection of 1-manifolds in $X$, 
hence a collection of embedded circles ($S^{1}$) or arcs $I_{j}$ with boundary $\partial I_{j} \subset F^{-1}(y \cup y')$. 
This gives a cobordism between $F^{-1}(y)$ and $F^{-1}(y')$ and it follows that the cardinality of $F^{-1}(y)$ is well-defined 
(mod 2). 

  For the same reasons, if $F: X \to Y$ is a proper Fredholm map of index $-1$ and $V$ is any properly embedded $1$-manifold 
in $Y$, with $V \cap \Ima F$ compact, then the $\bZ_{2}$ intersection number 
\be \label{intnum}
I_{\bZ_{2}}(F, V) \in \bZ_{2},
\ee
is defined as the cardinality of $F^{-1}(V')$, for any transversal approximation to $V'$. As above, this is well-defined. 

\medskip 

  Next we note a few standard elliptic regularity estimates to be used below. For the rest of the paper, we denote 
\be \label{notate}
\d_{0}^{*}X = \hat \cL_{X}g_{0}.
\ee
Modulo a factor of $2$, this is the trace-free part of the $L^{2}$ adjoint of the divergence operator $\d$. The operator 
$\d(\frac{1}{2N}\hat \cL_{(\cdot)}g_{0}) = \d (\frac{1}{2N}\d_{0}^{*})$ is formally self-adjoint and elliptic, and so has a discrete 
spectrum in $L^{2}$ with eigenvalues $\mu_{i} \in [0,\infty)$. Of course the eigenvectors with eigenvalue zero are exactly the 
conformal Killing fields. 
  
  Let $(\f, X)$ be a solution of the constraint equations 
\eqref{div}-\eqref{lich}. Elliptic regularity applied to the divergence constraint \eqref{div} gives the estimate 
\be \label{Xest}
|X|_{C^{1,\a}} \leq C |\f^{6}|_{L^{\infty}}|dH|_{L^{\infty}},
\ee
where the $C^{1,\a}$ and $L^{\infty}$ norms are with respect to $g_{0}$; the constant $C$ depends only on 
$M$, $\a$ and the representative $g_{0}$ for $[g] \in \cG$, cf.~\cite[Theorem 6.2.5]{Mor}. Also, observe that, 
modulo constants, $|X|_{C^{m,\a}} \leq |\d (\frac{1}{2N}\d_{0}^{*}X)|_{C^{m-2,\a}} \leq |\f^{6}dH|_{C^{m-2,\a}}$, 
so that 
\be \label{Xest2}
|X|_{C^{m,\a}} \leq C |\f^{6}|_{C^{m-2,\a}} |H|_{C^{m-1,\a}},
\ee
$m \geq 2$, with again $C$ depending only on $M$, $\a$ and $g_{0} \in \cG$. The estimates \eqref{Xest} and \eqref{Xest2} 
require that $(M, g_{0})$ has no conformal Killing fields; they hold for general $g_{0} \in Met^{m,\a}(M)$ if one assumes that 
$X$ is $L^{2}$ orthogonal to the space of conformal Killing fields on $(M, g_{0})$. However, the constant $C$ in \eqref{Xest} or 
\eqref{Xest2} will blow up on sequences $(g_{0})_{i} \in \cC'$ which converge to $g_{0} \in \cC \setminus \cC'$, i.e.~when 
$g_{0}$ has a conformal Killing field. This corresponds to the fact that the inverse operator to $\d (\frac{1}{2N}\d_{0}^{*})$ 
blows up on the space of eigenspaces with (arbitrarily) small eigenvalues. 

  For $(\f, X)$ as above, let $m = \max \f$ and let 
\be \label{renorm}
\bar \f = \frac{\f}{m}, \ \ \bar X = \frac{X}{m^{6}}.
\ee
Then \eqref{Xest2} gives 
\be \label{Xest3}
|\bar X|_{C^{m,\a}} \leq C |\bar \f^{6}|_{C^{m,\a}} |H|_{C^{m-1,\a}}.
\ee

\medskip 

  To conclude this section, note that the scale-invariance of the space of vacuum (Ricci-flat) Einstein metrics induces a scaling action on $\cC$. 
Thus if $(g, K) \in \cC$, then $(d^{4}g, d^{2}K) \in \cC$, for any $d > 0$. Under this action, 
\be \label{scale}
\f \to d \f, H \to d^{-2}H, X \to d^{4}X, \s \to d^{4}\s.
\ee

\section{Manifold structures}

  In this section, we study Banach manifold structures on the spaces $\cC'$ and $\cD'$ and analyse the map $\Pi$ in \eqref{Pi2} 
in more detail. 

  Let $\Lambda_{1}^{m-2,\a}(M)$ be the space of $1$-forms on $M$ with coefficients in $c^{m-2,\a}$ and consider the constraint 
map 
\be \label{Phi}
\Phi: Met^{m,\a}(M)\times S_{2}^{m-1,\a}(M) \to  c^{m-2,\a}(M)\times \Lambda_{1}^{m-2,\a}(M),
\ee
$$\Phi (g, K) = \left( \begin{array}{c} 
R_{g} - |K|^{2} + H^{2}\\ 
\d K + dH 
\end{array} \right).
$$
A simple inspection shows that the map $\Phi$ is well-defined and is a $C^{\infty}$ smooth map of Banach spaces, (or more 
precisely open domains of Banach spaces). If one fixes a volume form $\a$ and a representative $y = (g_{0}, \s, H) \in \cD^{m-1,\a}$ 
then it follows from the York decomposition as in \eqref{g}-\eqref{K} that the constraint map $\Phi$ takes the form 
\be \label{Phiy}
\Phi_{y}: c^{m,\a}(M) \times \chi^{m,\a}(M) \to c^{m-2,\a}(M)\times \Lambda_{1}^{m-2,\a}(M),
\ee
$$\Phi_{y}(\f, X) = \left( \begin{array}{c}
\f^{-5} \\ 
\f^{-6}
\end{array} \right) 
\cdot 
\left( \begin{array}{c} 
-\D \f + {\tfrac{1}{8}}R_{0}\f - {\tfrac{1}{8}}|\s + \frac{1}{2N}\d_{0}^{*}X|^{2}\f^{-7} + {\tfrac{1}{12}}H^{2}\f^{5} \\ 
\d (\frac{1}{2N}\d_{0}^{*}X) + {\tfrac{2}{3}}\f^{6}dH  
\end{array} \right). $$
Here $\chi^{m,\a}$ is the space of $c^{m,\a}$ vector fields on $M$ and we have used the notation from \eqref{notate}. 
Of course $(\f, X)$ depend on the choice of background metric $g_{0} \in [g]$ and volume form $\a$ while $(g, K)$ do not. 
Again $\Phi_{y}$ is a smooth map of Banach spaces. 

  It is well-known and easy to see that the system \eqref{Phiy} is a (non-linear, second order) elliptic system for the unknowns 
$(\f, X)$. For $(g_{0}, \s, H) \in \cD^{m-1,\a}$ and $N = \frac{1}{2}\frac{dv_{g_{0}}}{\a} \in c^{m,\a}$, the coefficients of the 
$2^{\rm nd}$ order derivatives of $(\f, X)$ are in 
$C^{m,\a}$, (in fact in $c^{m,\a} \subset C^{m,\a}$), the coefficients of the $1^{\rm st}$ order derivatives are in 
$C^{m-1,\a}$ while the coefficients of the $0$-order terms are in $C^{m-2,\a}$. Basic elliptic regularity estimates, 
cf.~\cite[Theorem 6.2.5]{Mor}, show that 
\be \label{ellest}
||(\f, X)||_{C^{m,\a}} \leq C[||D\Phi_{y}(\f, X)||_{C^{m-2,\a}} + ||(\f, X)||_{C^{0}}],
\ee
where $C$ depends only on the H\"older norms of the coefficients above. One has the same estimate for the formal $L^{2}$ adjoint 
of $D\Phi_{y}$.

  It follows from elliptic theory that the fiber map $\Phi_{y} = \Phi|_{\Phi^{-1}(y)}$ is Fredholm. It is 
for this reason that we choose to work with H\"older spaces. The elliptic estimate \eqref{ellest} does not hold for the 
non-linear map $\Phi_{y}$ when working with Sobolev spaces, cf.~again \cite[Theorem 6.2.5]{Mor}. It is unknown, (and 
probably not true) that $\Phi_{y}$ is Fredholm with respect to a Sobolev space topology. 

  The rows of \eqref{Phiy}, corresponding to the equations \eqref{div}-\eqref{lich}, are in general coupled, but are uncoupled 
and of Laplace type at leading order. Hence the Fredholm index of the map $\Phi_{y}$ is zero. The full constraint map 
$\Phi$ in \eqref{Phi} is an underdetermined elliptic operator; the linearization $D\Phi$ is semi-Fredholm, with finite dimensional 
cokernel but infinite dimensional kernel.  

\medskip 

  We first use the discussion above to describe the region where $\cC$ has the structure of a smooth Banach manifold. 
Given $(g, K)$, let $D_{(g,K)}\Phi$ be the linearization of $\Phi$ at $(g, K)$ and let $(D\Phi)^{*}$ denote the $L^{2}$ adjoint. 
Define the regular set 
\be \label{creg}
\cC^{reg} \subset \cC
\ee
to be the set of points $(g, K) \in \cC$ such that $\Ker (D_{(g,K)}\Phi)^{*} = 0$. We then have: 
\begin{theorem}
The space $\cC^{reg} \subset \cC^{m,\a}$ is a smooth separable Banach manifold. 
\end{theorem} 

{\bf Proof:} Naturally, the proof uses the implicit function theorem for Banach manifolds. 
To begin, one has the $L^{2}$ orthogonal splitting 
\be \label{Phisplit}
c^{m-2,\a}(M)\times \Lambda_{1}^{m-2,\a}(M) = \overline{\Ima D\Phi} \oplus \Ker (D\Phi)^{*}.  
\ee
Since $\Ker (D\Phi)^{*} = 0$ on $\cC^{reg}$, to apply the implicit function theorem, we need to 
show that $D\Phi$ has closed range and $\Ker D\Phi$ splits. As discussed above, the fiber map $D\Phi_{y}$ is of 
closed range with image of finite codimension. Let $S$ be a slice to $\Ima D\Phi_{y}$, so that $S$ is finite dimensional. 
Since $D\Phi$ has dense range, one may perturb $S$ slightly if necessary so that $S \subset \Ima D\Phi$ and choose a 
finite collection of ``vectors" $(h_{j}, \k_{j})$ in the domain of $D\Phi$ such that the collection $\{D\Phi(h_{j}, \k_{j})\}$ span $S$. 
It then follows easily that $D\Phi$ is of closed range. 

To see that $\Ker D\Phi$ splits, write 
$$T(Met^{m,\a}(M)\times S_{2}^{m-1,\a}(M)) = H \oplus V,$$
where $V =  \iota(T(c^{m,\a}(M) \times \chi^{m,\a}(M)))$; here $T(c^{m,\a}(M) \times \chi^{m,\a}(M))$ is the 
domain of $D\Phi_{y}$ and $\iota$ is the natural ``inclusion" map $\iota(\f, X) = (g, K)$ as in \eqref{div}-\eqref{lich}, given 
fixed data in $\cD$. The subspace $H$ corresponds to $T\cD$.  Since $\Ker D\Phi_{y}$ is finite dimensional and 
splits, one has 
$$V = \Ker D\Phi_{y}\oplus L,$$
where $L$ closed and of finite codimension in $V$. By construction, $\Ker D\Phi \cap L = 0$. We claim that 
$\Ker D\Phi \oplus L$ is of finite codimension in $T(Met^{m,\a}(M)\times S_{2}^{m-1,\a}(M))$. To see this, let $(g',K')$ 
be any variation of $(g, K)$ in $T(Met^{m,\a}(M)\times S_{2}^{m-1,\a}(M))$ and let $D\Phi(g',K') = w$. Recall that 
$\Ima D\Phi_{y}$ is of finite codimension. Thus if $w \in \Ima D\Phi_{y}$, there exists unique $(\f',X') \in L$ such that 
$D\Phi(g',K') - \iota(\f',X')) = 0$. This proves the claim. Since any space of finite codimension splits, it follows that 
$\Ker D\Phi$ splits. 

  This shows that $D\Phi$ is a submersion on $\cC^{reg}$ and the implicit function theorem (or regular 
value theorem) for Banach manifolds implies that the zero set 
$$\cC^{reg} = \Phi^{-1}(0)$$ 
is a smooth Banach manifold.  Thus $\cC^{reg}$ is an open Banach submanifold within $\cC$. 

{\endproof} 

\begin{remark}
{\rm Theorem 3.1 is discussed in detail and proved in the $C^{\infty}$ setting in \cite{FM2} and \cite{FMM}, by working in 
Sobolev spaces $H^{s}\times H^{s-1}$ and passing to the limit $s \to \infty$. However, as pointed out in \cite{B}, the proof of the 
manifold structure given in \cite{FM2} or \cite{FMM} does not hold when restricting to Sobolev spaces $H^{s}$ of finite differentiability. 
This issue is also discussed in detail in \cite{CD}. The proof the manifold theorem in \cite{B} holds in the low regularity space 
$H^{2}\times H^{1}$, but that argument does not generalize to $H^{s}$ spaces with $s > 2$, due to the failure of elliptic 
regularity estimates for the non-linear operator $\Phi$, as in \cite[Theorem 6.2.5]{Mor}. 

   The proof of Theorem 3.1 above, based on the conformal method, is somewhat different from the approaches above. 
}
\end{remark}

  Let 
\be \label{csing}
\cC^{sing}\subset \cC
\ee
denote the space of solutions with $\Ker (D_{(g,K)}\Phi)^{*} \neq 0$, so that $\cC^{sing} 
= \cC \setminus \cC^{reg}$ is closed and 
$$\cC = \cC^{reg} \cup \cC^{sing}.$$
The structure of $\cC$ near points $(g, K) \in \cC^{sing}$ 
has been analysed in detail in particular by Moncrief, Fischer and Marsden. To describe this, let $(\cM, g^{(4)})$ be the maximal 
vacuum Cauchy development of the initial data set $(M, g, K)$. Let $\nu$ be the unit (future-directed) time-like normal 
to $M$ in $(\cM, g^{(4)})$. Then by \cite{Mo}, $(N, Y) \in \Ker (D\Phi)^{*}$ if and only if the vector field $Z = N\nu + Y \in T\cM |_{M}$ 
extends to a space-time Killing field on $(\cM, g^{(4)})$. 

  Let $\cC^{cmc} \subset \cC$ denote the subspace of solutions where $H = const$. It is proved in \cite{FMM} that the space 
$\cC$ has cone-like singularities at the locus $\cC^{sing}\cap \cC^{cmc}$. Moreover, it is proved in \cite{FM1}, cf.~also \cite{BCS}, 
that for $H = const$, space-time Killing fields $Z$ are necessarily tangent to $M$ (so $N = 0$), except in the trivial case where 
$g_{0}$ is flat, $\f = const$ and $N = const$. Such $Z$ then give conformal Killing fields for $[g]$, which have been excluded 
in the definition above. 

  It follows that for the region $\cC^{cmc} \subset \cC'$, 
\be \label{cmcreg}
\cC^{cmc}\cap \cC^{sing} = \emptyset.
\ee
In particular, it follows that $\cC^{cmc}$ is a smooth Banach manifold. Moreover, it is well-known that the space of conformal 
classes with no conformal Killing field is open and dense in $\cG$, cf.~\cite{Eb} for instance. Hence, when $\cC^{cmc}$ is considered 
as a subset of $\cC$, one has 
\be \label{odcmc}
\overline{\cC^{reg} \cap \cC^{cmc}} = \cC^{cmc}.
\ee
The basic property \eqref{odcmc} is unknown however when $H \neq const$, cf.~\cite{BCS}. Some of the arguments in 
this work could be simplified if the analog of \eqref{odcmc} held, i.e.~if one had 
\be \label{regden}
\overline{\cC^{reg}} = \cC.
\ee

  For the work to follow we restrict the spaces $\cC^{reg}$ and $\cC^{sing}$ to $\cC'$, so work with 
the decomposition
\be \label{regsing'}
\cC' = \cC^{reg} \cup \cC^{sing}.
\ee

  We will discuss natural regularizations of $\cC^{sing}$ below, but first need to study the manifold structure 
of the target space $\cD$. As preceding \eqref{d'}, let $\cG' = (\cG')^{m,\a}$ be the space of 
$c^{m,\a}$ conformal classes which have no (non-zero) conformal Killing field. As in \S 1, we have the fibration 
\be \label{D'reg}
\pi: \cD' \to \cG'.
\ee
\begin{proposition} 
The space $\cD' = (\cD')^{m-1,\a}$ is a smooth separable Banach manifold and the projection map 
$\pi: \cD' \to \cG'$ is a smooth bundle map. 
\end{proposition} 

{\bf Proof:} This result is essentially well-known; the proof is based on the York decomposition \cite{Yo}, cf.~also 
\cite{FM2}, \cite{FM3}. To begin, consider the operator
$$\d_{0} = \d + {\tfrac{1}{3}}dtr : Met^{m,\a}(M)\times  S_{2}^{m-1,\a}(M) \to \Lambda^{m-1,\a}(M), \ \ (g, h) 
\to \d_{g}h + {\tfrac{1}{3}}dtr_{g}h.$$
Note that $\d_{0}$ is the $L^{2}$ adjoint of the conformal Killing operator $\hat \cL$, modulo a factor of $2$. We 
first claim that $\d_{0}$ is a submersion, so that the implicit function theorem implies that $\cZ = \d_{0}^{-1}(0)$ 
is a smooth separable Banach submanifold of $Met^{m,\a}(M)\times  S_{2}^{m-1,\a}(M)$. To see this, analogous to 
\eqref{Phisplit}, one has 
$$\Lambda_{1}^{m-1,\a} = \overline{\Ima \d_{0}} \oplus \Ker \d_{0}^{*}.$$
By assumption, $\Ker \d_{0}^{*} = 0$. To show that $\d_{0}$ has closed range, let $\w g$ be a metric 
in $(\cG')^{m+1,\a}$ sufficiently close to $g \in (\cG')^{m,\a}$ and let $\d_{0} = (\d_{0})_{g}$, $\w \d_{0} 
= (\d_{0})_{\w g}$. Consider the mapping
\be \label{dwd}
\d_{0} \w \d_{0}^{*}: \chi^{m+1,\a} \to \Lambda_{1}^{m-1,\a}.
\ee
This is an elliptic operator and so Fredholm for $\w g$ sufficiently near $g$. Also $\Ker \d_{0} \w \d_{0}^{*} = 0$, 
since this operator is a small perturbation of $\d_{0} \d_{0}^{*}$ which has no kernel by definition. It follows that 
$\d_{0}$ is of closed range and surjective. 

 To see that the kernel splits, given any $h \in T(S_{2}^{m,\a}(M))$, form $\d_{0}h$. The discussion above shows that 
for any such $\d_{0}h$, there is a unique $X$ such that $\d_{0}\w \d_{0}^{*}X = \d_{0}h$. Hence 
$h = (h - \w \d_{0}^{*}X) + \w \d_{0}^{*}X$ is the required splitting since $\w \d_{0}^{*}X \in S_{2}^{m,\a}(M)$. 
It thus follows from the implicit function theorem that $\cZ$ is a smooth separable Banach manifold. 

   Next, observe that the trace operator $tr: Met^{m,\a}(M)\times  S_{2}^{m-1,\a}(M) \to c^{m,\a}(M), \ \ (g, h) \to tr_{g}h$ 
is clearly a smooth submersion, so that $\cV = tr^{-1}(0)$ is a smooth separable Banach submanifold of 
$Met^{m,\a}(M)\times  S_{2}^{m-1,\a}(M)$. The intersection $\cZ \cap \cV$ is transverse, cf.~\cite{FM3} for instance, 
and hence the space of transverse-traceless tensors $\cZ \cap \cV$ is a smooth separable Banach submanifold of 
$Met^{m,\a}(M)\times  S_{2}^{m-1,\a}(M)$. Dividing out by the equivalence relation $(g, \s) \sim (\f^{4}g, \f^{-2}\s)$ gives 
a smooth separable Banach manifold structure to the quotient space $\cT'$. Crossing with the space $C^{m-1,\a}(M)$ of 
mean curvature functions, it follows again from the transversal intersection of $\cZ \cap \cV$ that  $\pi: \cD' \to 
\cG'$ is a smooth bundle projection. 

{\endproof} 

   Next we turn to the singular locus $\cC^{sing}$. As noted above, the structure of the singular locus $\cC^{sing}$ is not 
understood away from the the space of CMC solutions. For this reason, it is useful to regularize $\cC^{sing}$ by showing 
it can be naturally perturbed to a smooth manifold structure. To do this, consider the smooth map of Banach manifolds  
\be \label{Psi}
\Psi = (\Pi, \Phi): Met^{m,\a}(M)\times S_{2}^{m-1,\a}(M) \to \cD \times (c^{m-2,\a}\times \Lambda_{1}^{m-2,\a}),
\ee
$$\Psi(g, K) = (\Pi(g, K), \Phi(g, K)).$$
Note that $\cC = \Psi^{-1}(\ast, 0)$. Further the first factor $\Pi$ is trivially a surjective submersion onto $\cD$. The proof of 
Theorem 3.1 shows that $\Psi$ is a smooth Fredholm map, of Fredholm index zero. By the Sard-Smale theorem, the regular 
values of $\Psi$ are thus of second category. 

  Thus, given any $\e > 0$, there exist (many) values $(\mu, \xi) \in c^{m-2,\a}(M)\times \Lambda_{1}^{m-2,\a}(M)$ such that 
$|(\mu, \xi)| < \e$ and the inverse image 
\be \label{Ce}
\cC_{\e} := \Psi^{-1}(\ast, (\mu, \xi)),
\ee
is a smooth separable Banach submanifold of $Met^{m,\a}(M)\times S_{2}^{m-1,\a}(M)$. For any such regular value $(\mu, \xi)$, 
the space $\cC_{\e}$ is a smooth $\e$-approximation to the vacuum constraint space $\cC$. Choosing $\e = \e_{i} \to 0$ and 
corresponding $(\mu_{i}, \xi_{i}) \to (0, 0)$, the spaces $\cC_{\e_{i}}$ converge to $\cC$ in the following sense. If $\{(\mu_{i}, \xi_{i})\} 
\in \cC_{\e_{i}}$ is bounded in $C^{m-2,\a}$, then a subsequence converges to a limit $(\mu, \xi) \in \cC$. Conversely, any 
$(\mu, \xi) \in \cC$ is the limit of a bounded sequence $\{(\mu_{i}, \xi_{i})\} \in \cC_{\e_{i}}$. Of course any smooth compact 
subset of the regular set $\cC^{reg}$ is smoothly close to a domain in $\cC_{\e}$, for $\e$ sufficiently small.  

\begin{corollary}
For any regular value $\e$ as in \eqref{Ce}, the map  
\be \label{Pireg}
\Pi_{\e}: \cC_{\e} \to \cD
\ee
is a smooth Fredholm map of Banach manifolds of Fredholm index zero. 
\end{corollary}

{\bf Proof:} This follows directly from the proof of Theorem 3.1 and Proposition 3.3, using the well-known Fredholm alternative 
and the fact that $D\Phi_{y}$ is Fredholm of index zero. 
 
{\endproof}

  For the work to follow in \S 5, we will use an explicit parametrization of $\cD$, or more precisely a parametrization 
of the base space $\cG = \cG^{m,\a}$ of conformal classes. Let $\cY$ denote the space of Yamabe metrics 
in $Met^{m,\a}(M)$; thus $g_{0} \in \cY$ if and only if $g_{0}$ is a metric of constant scalar curvature and unit volume 
realizing the Yamabe invariant $Y[g_{0}]$ of $[g_{0}]$. (Yamabe metrics are always assumed to be minimizing metrics). 
It is proved in \cite{An} that there is an open-dense set 
\be \label{cy0}
\cY_{0} \subset \cY,
\ee
such that $g_{0} \in \cY_{0}$ is the unique (minimizing) Yamabe metric in its conformal class $[g_{0}]$. Moreover, there is a 
smooth bijection 
\be \label{cy1}
\iota: \cY_{0} \to \cU_{0} \subset \cG, 
\ee
onto an open-dense set $\cU_{0}$ in $\cG$. This gives $\cY_{0}$ the structure of a smooth Banach manifold, 
induced from the Banach manifold structure of $\cG$.  The space $\cU_{0}$ gives a natural parametrization for the 
space of equivalence classes $\cG_{0}$. Note that if $[g]$ admits a conformal Killing field which is 
not a Killing field for some $g_{0} \in \cY$, then $g_{0} \notin \cY_{0}$; namely the flow of $X$ then generates a 
1-parameter family of distinct Yamabe metrics. 

  By the solution to the Yamabe problem, the set of Yamabe metrics in a given conformal class $[g]$ is 
compact, away from the round conformal class $[g_{+1}]$ on $S^{3}$. Thus if $[g] \neq [g_{+1}]$ and $[g_{i}]$ is any 
sequence in $\cU_{0}$ with $[g_{i}] \to [g]$, then for the associated sequence of unique Yamabe metrics $(g_{0})_{i}$, 
there is a Yamabe metric $g_{0} \in [g]$ such that, in a subsequence,  
$$(g_{0})_{i} \to g_{0},$$
in $C^{m,\a}$. Further, for any other Yamabe metric $g_{0}' \in [g]$ there is a conformal factor $\psi'$ such that 
$g_{0}' = (\psi')^{4}g_{0}$. Thus for the conformal classes $\cG \setminus \cG_{0}$, the collection of conformal factors 
$\{\psi'\}$ in $[g]$ for (minimizing) Yamabe metrics is uniformly controlled in $C^{m,\a}$ for $[g] \neq [g_{+1}]$. 

  Note that this compactness and control of the conformal factors $\{\psi'\}$ does {\it not} hold for the case of the conformal 
class $[g_{+1}]$ of the round metric on $S^{3}$. This non-compactness is closely related to the Nirenberg problem and the 
Kazdan-Warner obstruction on $S^{3}$, cf.~\cite{HMM} for further details.

\section{Initial estimates.}

  In this section, we derive initial estimates on the behavior of solutions $(\f, X)$ of the constraint equations. These 
play an important role in the work to follow in \S 5. 

\medskip 

  We assume throughout this section (and the following) that $(\f, X)$ solve the constraint equations \eqref{div}-\eqref{lich} 
with volume form $\a$ fixed and with representative $(g_{0}, \s, H) \in \cD'$; in particular $[g]$ has no conformal Killing 
fields. Following this, we show that the same arguments extend to solutions of the constraint equations $\Phi(g, K) = (\mu, \xi)$, 
for any given fixed $(\mu, \xi) \in c^{m-2,\a}(M)\times \Lambda_{1}^{m-2,\a}(M)$. 

\begin{lemma} 
Suppose there is a constant $D < \infty$ such that 
\be \label{bb}
0 < D^{-1} \leq \inf \f \leq \sup \f \leq D < \infty.
\ee
Then there is a constant $C$, depending only on $D$ and the background data $(g_{0}, \s, H) \in \cD$ such that 
\be \label{sb}
|\f|_{C^{m,\a}} + |X|_{C^{m,\a}} \leq C.
\ee
\end{lemma} 

{\bf Proof:}  By \eqref{bb}, $\f$ and $\f^{-1}$ are bounded in $L^{\infty}$ by a fixed constant $D$. In particular, the right 
side of \eqref{div} is thus bounded in $L^{\infty}$, since $dH$ is bounded in $C^{m-2,\a}$. Elliptic regularity 
applied to the divergence constraint \eqref{div} as in \eqref{Xest} then gives  
$$|X|_{C^{1,\a}} \leq C,$$
since $\Ker \d (\frac{1}{2N}\d_{0}^{*}) = 0$. The right side of the Lichnerowicz equation \eqref{lich} is thus 
bounded in $C^{\a}$ and elliptic regularity applied to \eqref{lich} implies $\f$ is bounded in $C^{2,\a}$. 
In turn, this implies the right side of the divergence equation \eqref{div} is bounded in $C^{k,\a}$, $k = \min(2, m-2)$, 
and so elliptic regularity again implies $X$ is bounded in $C^{k+2,\a}$. Continuing this process inductively gives \eqref{sb}.

{\endproof}

\begin{proposition}
Let 
\be \label{supK}
\sup \f = M_{0}.
\ee
Then there is a constant $C < \infty$, depending only on $\max(1, M_{0})$, and the target data $(g_{0}, \s, H) \in \cD$, such that 
$$\sup \f \leq C \inf \f.$$
In particular, under an upper bound on $\f$, $\inf \f$ can approach $0$ only if $\sup \f$ approaches $0$. 
\end{proposition}

{\bf Proof:} The proof uses the well-known Moser iteration argument; we follow closely the description of this method 
in \cite[pp.194-198]{GT}.  All computations below are with respect to the background Yamabe metric $g_{0}$ 
with $R_{g_{0}} = Y[g]$ (or more generally a compact set of such metrics if $g_{0}$ is not unique). 

   To begin, from \eqref{lich} we have 
\be \label{mult}
-\f^{7+k}\D \f  = -{\tfrac{1}{8}}R_{0}\f^{k+8} + {\tfrac{1}{8}}|\s + {\tfrac{1}{2N}}\d_{0}^{*}X|^{2}\f^{k} - {\tfrac{1}{12}}H^{2}\f^{12+k}, 
\ee
cf.~again the notation \eqref{notate}. Integrating over $M$ and applying the divergence theorem gives 
\be \label{sob1}
-\int \f^{7+k}\D \f = \int \<d\f^{7+k}, d\f\> = (7+k)\int \f^{6+k}|d\f|^{2} = \frac{7+k}{(4+(k/2))^{2}}\int |d\f^{4+(k/2)}|^{2}.
\ee
Here and throughout the following, the integration over $M$ is with respect to the volume form of $(M, g_{0})$. Also, 
constants $c$, $C$, $c_{S}$, used below may change from line to line, or even inequality to inequality, but only depend on the 
target data $(M, g_{0}, \s, H)$ and $\a$. The Sobolev constant $c_{S}$ of $g_{0}$ is uniformly controlled, so that 
\be \label{Sob}
(\int \f^{6})^{1/3} \leq c_{S}\int (|d\f|^{2} + \f^{2}).
\ee
Applying this to $\f^{4+(k/2)}$ and using \eqref{sob1}, one obtains from \eqref{mult} that  
\be \label{est1}
\frac{7+k}{(4+k/2)^{2}}(\int \f^{24+3k})^{1/3} \leq C(\int |{\tfrac{1}{2N}}\d_{0}^{*}X|^{2}\f^{k} + \sup |\s|^{2}\int \f^{k} +   
|R_{0}|\int \f^{8+k} ),
\ee
for $k + 7 > 0$, where we have dropped the negative $H^{2}$ term. 

  If $k +7 < 0$, the sign changes; in this case we may drop the $\s$ and $\d_{0}^{*}X$ terms and obtain 
\be \label{est2}
\frac{|7+k|}{(4+k/2)^{2}}(\int \f^{24+3k})^{1/3} \leq C(|R_{0}|\int \f^{8+k} + \int H^{2}\f^{12+k} ),
\ee
provided $k + 8 \neq 0$. The case $k+8 = 0$ (the log case), will be considered later.

  We begin with the case $k + 7 > 0$, (the subsolution case). First, elliptic estimates for the divergence constraint 
\eqref{div} as in \eqref{Xest} imply that 
\be \label{hatXest}
|\d_{0}^{*}X|_{L^{4}} \leq c |X|_{L^{1,4}} \leq c |X|_{L^{2,2}} \leq c |\d ({\tfrac{1}{2N}}\d_{0}^{*}X)|_{L^{2}} \leq c(\int \f^{12})^{1/2},
\ee
where we have used the Sobolev inequality for the second inequality. By the H\"older inequality, this gives 
$$\int |\d_{0}^{*}X|^{2}\f^{k} \leq (\int |\d_{0}^{*}X|^{4})^{1/2}(\int \f^{2k})^{1/2} \leq c(\int \f^{2k})^{1/2} \int \f^{12}.$$
Inserting this in \eqref{est1} implies that 
$$\frac{1}{k}(\int \f^{24 + 3k})^{1/3} \leq c(\int \f^{2k})^{1/2} \int \f^{12} + c\int \f^{k} + c\int \f^{k+8},$$
where $c$ depends only on the target data $(g_{0}, \s, H)$. One may then iterate these inequalities, as in the usual Moser iteration, and 
starting with $k = 4$, obtain 
\be \label{est3}
\sup \f \leq C|\f|_{L^{12}} \leq C |\f|_{L^{2}},
\ee
where the last inequality follows from a standard interpolation inequality, \cite[p.146]{GT}. Again $C$ depends only on 
$(g_{0}, \s, H) \in \cD$. Note that the estimate \eqref{est3} does not require the assumption \eqref{supK}. Moreover, the 
bound \eqref{supK} only requires a bound on $H$ through the estimate of $\hat \cL_{X}g_{0}$ in \eqref{hatXest}.

  Next, as in \cite{GT}, consider the two cases $-1 < k+7 < 0$ and $k+7 < -1$. First, by \eqref{supK}, $H^{2}\f^{12+k} = 
H^{2}\f^{8+k}\f^{4} \leq H^{2}M_{0}^{4}\f^{8+k}$, so that \eqref{est2} implies that 
\be \label{est4}
\frac{|7+k|}{(4+k/2)^{2}}(\int \f^{3(8+k)})^{1/3} \leq C M_{0}^{4}\int \f^{8+k}.
\ee
Now first choose $k+8 = p \in (0,1)$ small. Then Moser iteration starting at $p$ and ending at $k+8 = 2$ shows that 
\be \label{est5}
\int \f^{2} \leq c(\int \f^{p})^{2/p},
\ee
for any $p > 0$ small, with $c = c(p, M_{0})$. 

  Next one may perform the same Moser iteration for $k + 8 < 0$ to obtain, for $p \in (0,1)$ as in \eqref{est5}, 
\be \label{est6}
(\int \f^{-p})^{-1/p} \leq c \inf \f,
\ee
with again $c = c(p, M_{0})$. To connect the estimates \eqref{est5} and \eqref{est6}, we claim that there is a constant 
$C = C(g_{0}, \s, H)$ and $p_{0} \in (0,1)$ such that 
\be \label{est7}
\int \f^{p_{0}} \int \f^{-p_{0}} \leq C. 
\ee

  For this, the $\log$ case, we return to the Lichnerowicz equation \eqref{lich} and write it as 
\be \label{log}
\f^{-1}\D \f = {\tfrac{1}{8}}R_{0} - {\tfrac{1}{8}}|\s + {\tfrac{1}{2N}}\d_{0}^{*}X|^{2}\f^{-8} + {\tfrac{1}{12}}H^{2}\f^{4}.
\ee
Integration, the divergence theorem and the estimate \eqref{supK}, together with the control on $R_{0}$ and $H$ imply that
$$\int |d \log \f|^{2} \leq CM_{0}^{4},$$
Next, still following \cite[p.198]{GT}, given any $p \in M$ and $r$ small, let $\eta = \eta(p,r)$ be a cutoff function satisfying $\eta = 1$ on 
the geodesic ball $B_{p}(r)$, $\eta = 0$ on $M \setminus B_{p}(2r)$ with $|d\eta| \leq C/r$. One has 
\be \label{loc}
\int_{B_{p}(r)}|d \log \f| \leq c(\int_{B_{p}(r)}|d \log \f|^{2})^{1/2}r^{3/2} \leq cr^{3/2}(\int_{B_{p}(2r)}|d \eta\log \f|^{2})^{1/2}.
\ee
Multiplying \eqref{log} by $\eta^{2}$ and integrating by parts in the same way, using also the Cauchy-Schwarz inequality and the 
scale change $r \to 2r$, gives 
for $r$ small, 
$$\int_{B_{p}(2r)}|d \log \f|^{2} \leq cr,$$
and hence by \eqref{loc}  
$$\int_{B_{p}(r)}|d \log \f| \leq Cr^{2}.$$
It then follows from the John-Nirenberg estimate \cite[p.166]{GT}, as in \cite[p.198]{GT}, that 
$$\int \f^{p_{0}} \int \f^{-p_{0}} \leq C,$$
for some $p_{0} \in (0,1)$, $C = C(M_{0})$, which proves \eqref{est7}. 

 Combining then \eqref{supK}, \eqref{est3}, \eqref{est5}-\eqref{est7} shows that 
$$1 = \sup \f \leq C(\int \f^{p_{0}})^{1/p_{0}} \leq C(\int \f^{-p_{0}})^{-1/p_{0}} \leq C\inf \f,$$
which proves the result. 

{\endproof}

  Proposition 4.2 shows that an upper bound on $\sup \f$ gives control of the Harnack constant 
\be \label{Harn}
C_{Har}(\f) = \frac{\sup \f}{\inf \f},
\ee
of $\f$, given control of the target data in $\cD$. As an application of Proposition 4.2, we prove the following: 

\begin{proposition} 
Continuing under the assumption \eqref{supK}, suppose there is a constant $s_{0} > 0$ such that $\inf |\s| \geq s_{0} > 0$. 
Then there is a constant $\k_{0} > 0$, depending only on $M_{0}$, $s_{0}$, $\a$ and $(g_{0}, \s, H) \in \cD$, such that 
\be \label{inf1}
\inf \f \geq \k_{0} > 0.
\ee
Moreover, if $Y(g) \leq -Y_{0} < 0$, then 
\be \label{inf2}
\inf \f \geq \k_{0} > 0,
\ee
where $\k_{0}$ depends only on $M_{0}$, $Y_{0}$ and $(g_{0}, \s, H) \in \cD$. 
\end{proposition}

{\bf Proof:} To prove \eqref{inf1}, by Proposition 4.2 it suffices to obtain a lower bound on $m_{0} = \sup \f$. Namely if $m_{0} \leq 1$ 
then the bound on $C_{Har}$ from Proposition 4.2 shows that a lower bound on $\inf \f$ and $\sup \f$ are equivalent. 
Now integrating the Lichnerowicz equation \eqref{lich} over $(M, g_{0})$ gives 
$$\int_{M}|\s + {\tfrac{1}{2N}}\d_{0}^{*}X|^{2}\f^{-7} \leq {\tfrac{1}{8}}|R_{0}| \int_{M}\f + {\tfrac{1}{12}}\sup H^{2}\int_{M}\f^{5} \leq c m_{0},$$
for a fixed constant $c$. We assume here without loss of generality that $m_{0} \leq 1$. Since $\f^{-7} \geq m_{0}^{-7}$, it follows that 
$$m_{0}^{-7}\int_{M}|\s|^{2} \leq m_{0}^{-7}\int_{M}|\s + {\tfrac{1}{2N}}\d_{0}^{*}X|^{2} \leq \int_{M}|\s + {\tfrac{1}{2N}}\d_{0}^{*}X|^{2} \f^{-7} \leq cm_{0},$$
so that 
$$\int_{M}|\s|^{2} \leq c m_{0}^{8}.$$ 
Now $|\s|^{2}$ is controlled in $c^{m-1,\a}$ and so the bound $\inf |\s| := |\s|(p) \geq s_{0} > 0$ implies there is a fixed $r_{0}$ such that 
$|\s|(x) \geq s_{0}/2$ for all $x \in B_{p}(r_{0})$. It follows that 
$$r_{0}^{4}s_{0}^{2} \leq \int_{M}|\s|^{2} \leq c m_{0}^{8}.$$
This gives a lower bound for $m_{0}$ in terms of $\s$ and $s_{0}$, which thus proves \eqref{inf1}. 

 For \eqref{inf2}, evaluating the Lichnerowicz equation \eqref{lich} at a point $p$ realizing $\min \f = \inf \f$ gives 
$$0 \leq {\tfrac{1}{12}}H^{2}(p)(\inf \f)^{5} + {\tfrac{1}{8}}R_{0} \inf \f,$$
(regardless of the behavior of $\s$ and $\d_{0}^{*}X$). Recall that $R_{0}$ is the Yamabe constant $Y[g]$ of $[g]$. 
If $R_{0} < 0$, then $H^{2}(p)(\inf \f)^{4} \geq \frac{3}{2}|R_{0}|$, which proves \eqref{inf2}. 

{\endproof}

\begin{remark} 
{\rm When $Y[g] > 0$, simple examples show that \eqref{inf1} is not true without the assumption on $\s$. Thus, suppose 
$g_{0}$ is the standard product metric on $S^{1}(1)\times S^{2}(1)$, so that $R_{g_{0}} = 2$. 
Choose 
$$\s = \k(-d\theta^{2}+ {\tfrac{1}{2}}g_{S^{2}(1)}),$$
for some constant $\k$. The form $\s$ is transverse-traceless with respect to $g_{0}$ and has constant norm $|\s|^{2} = 
\frac{3}{2}\k^{2}$. Let also $H = c$, an arbitrary constant. Then the divergence constraint \eqref{div} is satisfied by setting $X = 0$ 
while the Lichnerowicz equation \eqref{lich} holds if $\f = \e = const$ and 
$$0 = 2\e - |\s|^{2}\e^{-7} + {\tfrac{2}{3}}H^{2}\e^{5}.$$
This holds by choosing $\k$ so that $\frac{3}{2}\k^{2} = 2\e^{8} + \frac{2}{3}H^{2}\e^{12}$. 
 
   This example shows that one may have $Y(g) > 0$ with $H$ an arbitrary constant, with $\f \to 0$ uniformly as 
$\s \to 0$ uniformly. 
}
\end{remark}

\begin{remark} 
{\rm Similarly, there are numerous examples of curves $(g_{t}, K_{t})$, $t \in [0, \infty)$ with $Y(g_{t}) \leq -c < 0$ where 
$H_{t} \to 0$, $\s_{t} \to 0$ and $\f_{t} \to \infty$ pointwise as $t \to \infty$. The simplest examples are the 
Milne universe or hyperbolic cone metric 
$$g^{(4)} = -dt^{2} + t^{2}g_{-1},$$
where $(M, g_{-1})$ is a hyperbolic $3$-manifold. This is a flat (and hence Ricci-flat) Lorentz metric on $\bR^{+}\times M$. 
One easily sees that on the slices $M = M_{t} = \{t = const\}$, $\s = 0$, $\f_{t} = \sqrt{t}\to \infty$ and $H_{t} = 
\frac{3}{t} \to 0$ as $t \to \infty$. 

 Similar behavior occurs in the long-time future behavior of vacuum space-times near the flat hyperbolic cone space-time 
 by the work of Andersson-Moncrief \cite{AM}, as well as in the $U(1)$-symmetric space-times of  Choquet-Bruhat-Moncrief 
\cite{CBM}. 

}
\end{remark}

\begin{remark}
{\rm It is easy to see that all of the results of this section hold for $(\f, X) \in \cC_{\e}$, i.e.~for $(\f, X)$ satisfying 
the non-vacuum constraint equations $\Phi(g, K) = (\mu, \xi)$, for fixed $(\mu, \xi) \in c^{m-2,\a}(M)\times \Lambda_{1}^{m-2,\a}(M)$ 
provided (say) $\mu \geq 0$. Namely, the divergence and Lichnerowicz equations \eqref{div}-\eqref{lich} then take the form 
\be \label{divX}
\d ({\tfrac{1}{2N}}\d_{0}^{*}X) = (-{\tfrac{2}{3}}dH + \xi)\f^{6},
\ee
\be \label{lichN}
8\D \f = R_{0}\f - |\s + {\tfrac{1}{2N}}\d_{0}^{*}X|^{2}\f^{-7} + ({\tfrac{2}{3}}H^{2} + \mu)\f^{5}.
\ee  
Since $\mu \geq 0$, a brief inspection shows that all the results of this section hold as before, with constants depending 
only on the $c^{m-2,\a}(M)\times \Lambda_{1}^{m-2,\a}(M)$ norm of $(\mu, \xi)$. 

}
\end{remark}

  Summarizing briefly, the results of this section show that, given control on the target data in $\cD'$, there is no 
degeneration of the fiber data $(\f, X)$ when $Y(g) < 0$ is bounded away from $0$, or when $Y(g) \geq 0$ and $\s$ 
bounded away from 0 at some point, {\it provided} one has a $\sup$ bound on $\f$. This shows that such control of 
the target data in $\cD'$ and control of $\sup \f$ implies control of the fiber data ($\f, X)$. Note also that by 
\eqref{est3}, 
$$\sup \f \leq C |\f|_{L^{p}},$$
for any $p \geq 2$ (say), so that it suffices to obtain uniform control on the $L^{p}$ norm of $\f$.

\section{Slicings and Proper Maps.}

 In this section, we consider natural slicings of the space $\cC$ and analyse the properness of the map $\Pi$ when 
restricted to the individual slices. 

  While there are many natural slicings one might consider, we work with the slicing discussed in \S 1 given by the $L^{p}$ norm 
of $\f$. Let $\hat \cD \subset \cD'$ denote the space over the conformal classes $\cU_{0}$ which have a unique 
Yamabe metric, cf.~\S 3. Let $\hat \cC = \Pi^{-1}(\hat \cD)$ and note that  $\hat \cC$ is open and dense in $\cC'$. We recall 
again that $\a$ is fixed (but arbitrary) but the notation $\Pi$ is used for $\Pi_{\a}$. For any 
$p \geq 1$, the functional 
$$F_{p}(g, K) = \int_{M}\f^{p} dv_{g_{0}}: \hat \cC \to \bR^{+},$$
is well-defined and continuous. On the regular set $\cC^{reg} \subset \hat \cC$ where $\cC'$ is a smooth Banach manifold, 
the functional $F_{p}$ is smooth. Since the space of minimizing Yamabe metrics in a fixed conformal class $[g] \in \cG'$ is compact, 
$F_{p}$ extends to a bounded function on $\cC'$, multivalued in $\cC' \setminus \hat \cC$. Using a $C^{0}$ partition of unity, 
the function $F_{p}$ may be mollified in a neighborhood of $\cC'\setminus \hat \cC$ to give a continuous function 
$$F_{p}(g, K) = \int_{M}\f^{p} dv_{g_{0}}: \cC' \to \bR^{+}.$$

  The most natural choice geometrically for $p$ is $p = 6$, giving $F(g, K) = vol_{g}M$. However for the purposes below 
we will need to choose $p > 6$ and so for convenience choose $p = 8$: 
\be \label{f8}
\begin{array}{cc}
F = F_{8}: \cC' \to \bR^{+}, \\
F(g,K) = \int_{M}\f^{8}dv_{g_{0}}.
\end{array}
\ee

  Let 
$$\cC^{\o} = \{(g, K) \in \cC': F(g,K) = \o\},$$
so that $\cC'$ is foliated by the level sets $\cC^{\o}$ of $F$. 
Recall from \eqref{scale} that $\cC'$ is invariant under the scaling $(g, K) \to (d^{4}g, d^{2} K)$. Under this scaling one has 
$$\cC^{\o} \to \cC^{d^{8}\o},$$
so that all level sets $\cC^{\o}$ are homeomorphic and one has a global splitting 
$$\cC' = \cC^{\o} \times \bR^{+}.$$ 
In the regular region where $\cC'$ is a smooth manifold, the level sets $\cC^{\o}\cap \cC^{reg}$ are 
smooth hypersurfaces of $\cC^{reg}$ and the splitting 
$$\cC^{reg}= (\cC^{\o}\cap \cC^{reg}) \times \bR^{+},$$
is smooth.  

  Now consider the mapping 
\be \label{piw}
\begin{array}{cc}
\Pi^{\o}: \cC^{\o} \to \cD', \\
\Pi^{\o} = ([g, \s], H).
\end{array}
\ee
By the results of \S 3, in the region $\cC^{\o}\cap \cC^{reg}$ the map $\Pi^{\o}$ is smooth and Fredholm, of Fredholm 
index $-1$, for any $\o$. Similarly, for any regular value $\e > 0$, the map 
\be \label{piwe}
\begin{array}{cc}
\Pi_{\e}^{\o}: \cC_{\e}^{\o} \to \cD', \\
\Pi_{\e}^{\o} = ([g, \s], H).
\end{array}
\ee
is a smooth Fredholm map of Banach manifolds, with Fredholm index $-1$. 

  The first main result of this section corresponds to Theorem 1.1: 
  
\begin{theorem} 
For any $\o > 0$, the map $\Pi^{\o}$ is continuous and proper. Similarly, for any $\o > 0$ and $\e > 0$, 
the map $\Pi_{\e}^{\o}$ is smooth and proper. 
\end{theorem}

{\bf Proof:} This follows easily from the results in \S 4, using the Yamabe representatives $g_{0}$ for conformal 
classes $[g]$, as discussed above. Namely, Proposition 4.2 gives first a $\sup$ bound on 
$\f$, cf.~\eqref{est3} and therefore also an $\inf$ bound. The result then follows from Lemma 4.1. The second 
statement follows from Remark 4.6. 

{\endproof} 

 We note that the proof of properness does not use or require a manifold structure for the domain space $\cC^{\o}$. 
As noted in \S 1, the maps $\Pi$ or $\Pi^{\o}$ are not likely to be proper over points $[g]$ which contain a conformal Killing field. 

  Next we analyse the intersection properties of the image $\Ima \Pi^{\o}$ with natural choices of lines in $\cD'$. 
In this work, we consider lines of the form $\{\s\} = \{\l \s\}$, $\l \in \bR$, $\s \neq 0$. One might also consider 
lines of the form $\l H$, $H + \l$, $\s + \l \s_{0}$ for a fixed $\s_{0}$, and so on, but this will not be carried out here. 

\begin{proposition} 
For any given $\o > 0$ and any line $L_{\s} = ([g, \l \s], H)$, $\l \in \bR$, (with $([g, \s], H)$ fixed), the intersection 
\be \label{compint}
\Ima \Pi^{\o} \cap L_{\s}
\ee
is compact in $\cD'$. Equivalently, for $\ell_{\s} = \Pi^{-1}(L_{\s})$, 
\be \label{compint2}
\ell_{\s} \cap \cC^{\o},
\ee
is compact in $\cC'$. 
\end{proposition}

{\bf Proof:} By Theorem 5.1, \eqref{compint} and \eqref{compint2} are equivalent and it suffices to prove that 
\be \label{ul}
|\l| \leq \Lambda_{0} < \infty,
\ee
for some $\Lambda_{0}$ depending only on $([g, \s], H)$, (and $\a$). Without loss of generality, 
we may assume $|\s|_{C^{m-1,\a}} = 1$. 

  To start, the divergence constraint \eqref{div} does not involve $\l$. Using elliptic regularity, the $L^{8}$ 
bound on $\f$ and control on $H$ imply that $X$ is uniformly bounded or controlled in $L^{2,4/3}$, so that $DX$ is 
uniformly controlled in $L^{1,4/3} \supset L^{12/5}$, by Sobolev embedding in dimension $3$. In particular, $\d_{0}^{*}X$ 
is uniformly bounded in $L^{2}$.  Now the Lichnerowicz equation \eqref{lich} gives  
$$8\D \f = R_{0}\f - |\l \s + {\tfrac{1}{2N}}\d_{0}^{*}X|^{2}\f^{-7} + {\tfrac{2}{3}}H^{2}\f^{5}.$$
Integrating this over $M$ and using the $L^{8}$ bound on $\f$ and the control on $R_{0}$ and $H$, follows that 
$$\int_{M}|\l \s + {\tfrac{1}{2N}}\d_{0}^{*}X|^{2}\f^{-7} \leq C,$$
for some fixed constant $C$. 

  Since $|\s|_{C^{m-1,\a}} = 1$, there is an open set $U \subset M$ and constants $d > 0, v_{0} > 0$ (depending only on 
$g_{0}$, $\s$) such that $|\s| \geq d$ pointwise on $U$ with $vol U \geq v_{0}$. One has $vol \, U = \int_{U}1 = 
\int_{U}\f \f^{-1} \leq (\int_{U}\f^{-7})^{1/7}(\int_{U}\f^{7/6})^{6/7}$ and so 
$$\int_{U}\f^{-7} \geq (vol \, U)^{7} / (\int \f^{7/6})^{6}.$$
We have $|\f|_{L^{7/6}} \leq |\f|_{L^{8}}$, and hence 
$$\int_{U}\f^{-7} \geq d',$$
with $d'$ depending only on $d$, $v_{0}$ and $\o$. Since 
$$\inf_{U}|\l \s + {\tfrac{1}{2N}}\d_{0}^{*}X|^{2}\int_{U}\f ^{-7} \leq  \int_{U}|\l \s + {\tfrac{1}{2N}}\d_{0}^{*}X|^{2}\f^{-7},$$
it follows that  
\be \label{infd}
\inf_{U}|\l \s + {\tfrac{1}{2N}}\d_{0}^{*}X|^{2} \leq C.
\ee
However $|\l \s| \geq d\l$ pointwise everywhere on $U$ so that \eqref{infd} implies (for $\l$ sufficiently large) that 
$|\frac{1}{2N}\d_{0}^{*}X| \geq \frac{d}{2}\l$ pointwise on $U$. Since the $L^{2}$ norm of $\frac{1}{2N}\d_{0}^{*}X$ on $M$ 
is uniformly bounded, this proves \eqref{ul}. 

{\endproof} 

  It is clear that Proposition 5.2 also holds for the mapping $\Pi_{\e}^{\o}$ in \eqref{piwe}. 
  
   We will calculate the intersection number of $\Pi^{\o}$ with $L_{\s}$ generally in \S 5, but it is worthwhile to discuss the behavior 
of $\Pi^{\o}$ on the space $\cC_{cmc}$ of CMC solutions, i.e. 
$$\Pi_{cmc}^{\o}: \cC_{cmc}^{\o} \to \cD_{cmc}.$$ 
(The script $cmc$ has been lowered for notational convenience). As discussed in \S 2, in this case, $\Pi_{cmc}$ is a smooth 
diffeomorphism onto its image $\cD_{cmc}^{ex}$ and $\Pi_{cmc}^{\o}$ is a smooth embedding of codimension $1$. 
For each $([g, \s], H) \in \cD_{ex}$, there exists a unique solution $\f = (\f, 0)$ of the Lichnerowicz equation \eqref{lich}. 
(The divergence constraint \eqref{div} is trivially uniquely satisfied by setting $X = 0$). 

  Thus fix the line $L_{\s} = ([g, \l \s], H)$, $H = const$. For each $\l$ there is a unique solution $\f = \f(\l)$, so that 
$\o = \o(\f)$ is then a well-defined smooth function of $\l$; 
\be \label{ol}
\o = \o(\l): \bR \to \bR^{+}.
\ee

\begin{proposition} 
The function $\o$ in \eqref{ol} is a proper map $\o: \bR \to \bR^{+}$ and 
\be \label{degw0}
deg \, \o = 0.
\ee
The $\bZ_{2}$ intersection number is given by  
\be \label{int00}
I_{\bZ_{2}}(\Pi_{cmc}^{\o}, L_{\s}) = 0 \ (mod \ 2).
\ee
\end{proposition}

{\bf Proof:}  We recall that the degree of a smooth map $\o: \bR \to \bR^{+}$ is the number of solutions of $\o(\l) = \o_{0}$, for 
$\o_{0}$ a regular value of $\o$, counted with signed multiplicity according to whether the derivative $\o'$ is positive or 
negative at a regular point in $\o^{-1}(\o_{0})$. 

  As noted in \S 1, CMC solutions $(\f, 0)$ of the constraint equations \eqref{div}-\eqref{lich} naturally come in pairs in that 
$(\f, 0)$ is a solution with data $(]g, \l \s], H)$ if and only if it is also a solution with data $([g, -\l \s], H)$. 
This implies immediately that 
$$\o(\pm \l) = \o(\l).$$
It is easy to see that the sign of the derivative $\o'$ changes on passing from $\l$ to $-\l$, which gives \eqref{degw0} and 
\eqref{int00}. 

{\endproof} 

  It is worthwhile to describe the behavior of $\o$ in more detail in the three cases $Y[g] > 0$, $Y[g] = 0$ and $Y[g] < 0$. 
The results below follow from a simple analysis of the Lichnerowicz equation \eqref{lich}. 

\medskip 
  
  {\bf (i).} ($Y[g] > 0$). The inverse image $\ell_{\s} = \Pi^{-1}([g, \l \s], H)$, $H = const$, consists of two distinct components $\ell_{\s}^{\pm}$
with $\ell_{\s}^{-}$ parametrized by $\l \in (-\infty, 0)$, $\ell_{\s}^{+}$ parametrized by $\l \in (0, \infty)$ and satisfying 
$$\o \to 0 \ {\rm as} \ |\l| \to 0, \ \ \o \to \infty \ {\rm as} \ |\l| \to \infty.$$

  {\bf (ii).} ($Y[g] = 0$). If $H = const \neq 0$, then there is a minimal value $\l_{0} = \l_{0}(H) > 0$ such that $\ell_{\s}$ has two distinct 
components $\ell_{\s}^{\pm}$ parametrized by $\l \in (-\infty, -\l_{0}) \cup (\l_{0}, \infty)$ with 
$$\o \to 0 \ {\rm as} \ |\l| \to \l_{0}, \ \ \o \to \infty \ {\rm as} \ |\l| \to \infty.$$
If $H = 0$, the only solution is over $\l = 0$ with $\f = const$, so that the parametrization by $\l$ breaks down. 

{\bf (iii).} ($Y[g] < 0$). Then $H \neq 0$ and $\ell_{\s}$ is a single connected curve with $\l \in (-\infty, \infty)$ and with $\o$ achieving 
a minimum value $\o_{0}$ at $\l = 0$, so that 
$$\o(\l) \geq \o_{0},$$
with $\o(0) = \o_{0} > 0$. As $H \to 0$, $\o_{0} \to \infty$. 
 
\medskip  
 
  The following lower bound on $\l$ when $Y[g] > 0$ will be important in \S 6. Recall that $\{g, \s\}$ denotes the 
equivalence class $[g, \s] \sim [g, \l \s]$. 
\begin{proposition}  
If $Y[g] > 0$, there are constants $\o_{0} > 0$, $c_{0} < \infty$, depending only on $(\{g, \s\}, H)$ such that for any solution 
$(\f, X) \in \cC^{\o}$ over $(\{g, \s\}, H)$ with $\o \leq \o_{0}$, one has   
\be \label{ll0}
\l \geq c_{0}\o^{1/2} > 0.
\ee
\end{proposition}

{\bf Proof:}  Write the Lichnerowicz equation \eqref{lich} as 
\be \label{main}
{\tfrac{2}{3}}H^{2}\f^{12} + R_{0}\f^{8} - 8\f^{7}\D \f = |\l \s + {\tfrac{1}{2N}}\d_{0}^{*}X|^{2} \leq \l^{2}|\s|^{2} + 
({\tfrac{1}{2N}})^{2}|\d_{0}^{*}X|^{2}. 
\ee
At a point $p$ realizing $\sup \f$, $-\D \f \geq 0$, so that all terms on the left in \eqref{main} are non-negative. 
By the control on $(\f, X)$ from Theorem 5.1 and Proposition 5.2, together with the estimates \eqref{Xest} and 
\eqref{est3}, one has $|\d_{0}^{*}X|^{2} = O(\sup \f^{12}) = O(\o^{3/2})$. If $\sup \f$, or equivalently $\o$, is sufficiently 
small, the dominant term on the left in \eqref{main} is the $R_{0}\f^{8}$ term, (or the possibly larger $-\f^{7}\D \f$ term). 
Since $|\d_{0}^{*}X|^{2}$ or $(\frac{1}{2N})^{2}|\d_{0}^{2}X|^{2}$, is much smaller than $O(\sup \f^{8})$, it follows that 
\be \label{lw}
\l^{2} \geq c_{0}\o,
\ee
for $\o$ sufficiently small. This proves \eqref{ll0} on $\cC^{\o}$. 

 {\endproof}

  When $Y[g] < 0$, the analog of Proposition 5.4 becomes vacuous. 
\begin{proposition}
If $Y[g] < 0$. there exists $\o_{0} > 0$, depending only on $(\{g, \s\}, H)$ such that there are no solutions of the 
constraint equations \eqref{div}-\eqref{lich} in $\cC^{\o}$, for $\o \leq \o_{0}$. 
\end{proposition}

{\bf Proof:} If $(\f, X)$ solve the constraint equations, then evaluating the Lichnerowicz equation \eqref{lich} at a point 
$p$ realizing $\min \f$ gives 
\be \label{neg}
0 \leq R_{0}\f - |\l \s + {\tfrac{1}{2N}}\d_{0}^{*}X|^{2}\f^{-7} + H^{2}\f^{5}.
\ee
By Theorem 5.1, $\l$ is bounded above and hence (as discussed in the proof of Theorem 5.1) all solutions $(\f, X)$ of the 
constraints are uniformly controlled by the target data $(\{g], \s\}, H)$. Hence if $\o$ is sufficiently small and $R_{0} < 0$, 
$$H^{2}\f^{5} < < |R_{0}|\f$$
when evaluated at $p$. This contradicts \eqref{neg}. 

{\endproof}

\begin{remark} 
{\rm The behavior in the transition or borderline region $Y[g] = 0$ between $Y[g] < 0$ and $Y[g] > 0$ where $\o$ is small 
is rather subtle. We will see in \S 6, cf.~Theorem 6.3, that solutions always exist with $\l$, $\o$ small when $Y[g] > 0$, 
so the transition $Y[g] < 0$ to $Y[g] > 0$ is from existence to non-existence of solutions (with sufficiently small $\o$). 
This is discussed in more detail in Remark 6.6. 

  We make one further remark in the $Y[g] = 0$, i.e.~$R_{0} = 0$ case. Pairing the Lichnerowicz equation \eqref{lich} with $\f^{7}$ 
and integrating over $M$ gives 
$$\int_{M}{\tfrac{2}{3}}H^{2}\f^{12} + {\tfrac{7}{2}}|d\f^{4}|^{2} = \int_{M} |\l \s|^{2}  + ({\tfrac{1}{2N}})^{2}|\d_{0}^{*}X|^{2}.$$ 
By the Sobolev inequality \eqref{Sob}, this implies 
\be \label{main2}
\int_{M}{\tfrac{2}{3}}H^{2}\f^{12} + c\int_{M}(\f^{4} - \bar \f^{4})^{6})^{1/3} \leq \int_{M} |\l \s|^{2}  + ({\tfrac{1}{2N}})^{2}|\d_{0}^{*}X|^{2}, 
\ee
where $\bar \f^{4} = \int_{M}\f^{4}$ and $c > 0$. As before, by the control on $(\f, X)$ from Theorem 5.1, Proposition 5.2, 
and the estimates \eqref{Xest}, \eqref{est3}, one 
has $|\d^{*}X|^{2} = O(\sup \f^{12}) = O(\inf \f^{12}) = O(\o^{3/2})$. Hence, if $\int_{M}(\f^{4} - \bar \f^{4})^{6})^{1/3} >> 
O(\f^{12})$, one obtains a bound on $\l$ away from $0$ analogous to \eqref{ll0}, i.e. 
$$\l^{2} \geq C \int_{M}(\f^{4} - \bar \f^{4})^{6})^{1/3},$$
with $C$ large. If there is no such estimate, then $\l^{2} \leq O(\sup \f^{12})$ and moreover 
$$\int_{M}(\f^{4} - \bar \f^{4})^{6})^{1/3} \leq c \f^{12}.$$
This means that $\f$ is very close to being a constant function relative to its size when $\o$ is small. This may be useful 
for future studies. 

}
\end{remark}

\begin{remark}
{\rm Just as in Remark 4.6, the results in the section do not require restriction to the vacuum constraint equations. 
All results hold for pairs $(g, K) \in \cC_{\e}$ for which $\Phi(g, K) = (\mu, \xi)$ with $(\mu, \xi)$ controlled in $c^{m-2,\a}(M)\times 
\Lambda_{1}^{m-2,\a}$ with and $\mu \geq 0$. 

}
\end{remark}

\section{Existence and Non-existence results}

  In this section, we combine the work in previous sections to prove Theorems 1.2 and 1.3 and several related results.  

\medskip 

   Let $\cC_{\e}$ be an $\e$-regularization of $\cC'$ as in \S 3 and consider the map 
\be \label{Piwe2}
\Pi_{\e}^{\o}: \cC_{\e}^{\o} \to \cD'.
\ee
This is a smooth Fredholm map between separable Banach manifolds and we apply the global analysis methods of Smale 
discussed in \S 2. Note that $\Pi_{\e}^{\o}$ is of Fredholm index $-1$, (since the map $\Pi_{\e}: \cC_{\e} \to \cD'$ 
is of Fredholm index $0$). By Theorem 5.1, $\Pi_{\e}^{\o}$ is a proper Fredholm map. We consider then natural properly 
embedded $1$-manifolds $V \subset \cD'$ transverse to $\Pi_{\e}^{\o}$ and consider the corresponding intersection 
properties. 

   As in \S 5, we choose $V = L_{\s} = ([g, \l \s], H)$, $\l \in \bR$, with $([g], H)$ arbitrary. 
If $\cP'$ denotes the projectivization of $\cD'$ (i.e.~one projectivizes the fibers $\s \sim \l \s$), the map 
\be \label{wPiwe2}
\w \Pi_{\e}^{\o}: \cC_{\e}^{\o} \to \cP',
\ee
is a Fredholm map of Fredholm index $0$ away from solutions $(\f, X)$ which have data with $\l = 0$.  
Note that $\w \Pi_{\e}^{\o}$ is not defined on $(\f, X)$ with $\l(\f, X) = 0$; this is the reason for working with the 
intersection behavior of lines with $\Pi_{\e}^{\o}$ in place of the degree behavior of $\w \Pi_{\e}^{\o}$.  

   The following result is a more precise version of Theorem 1.2. 

\begin{theorem}
For any regular value $\e > 0$, one has 
\be \label{deg10}
I_{\bZ_{2}}(\Pi_{\e}^{\o}, L_{\s}) = 0. 
\ee
\end{theorem}

{\bf Proof:} By the results of \S 5, the intersection number $I_{\bZ_{2}}$ defined in \eqref{intnum} is well-defined and independent 
of $\o$ and the data $(\{g, \s\}, H)$. When $Y[g] < 0$, Proposition 5.5 (and Remark 5.7) shows 
that $I_{\bZ_{2}} = 0$ for $\o$ sufficiently small. Hence \eqref{deg10} holds for all $\o$, $\e$, and all data $(\{g, \s\}, H)$. 

{\endproof} 

\begin{remark}
{\rm It seems very likely that Theorem 6.1 and \eqref{deg10} hold also for $\e = 0$, i.e.~on $\cC'$. Namely, for a given regular value 
$(\mu, \xi)$, consider the region 
$$\cC(\e_{0}) = \{(g, K): \Phi(g, K) = t(\mu, \xi), |t| < \e_{0}\},$$
so that $\cC(\e_{0})$ is the union of the spaces $\cC_{t}$, $|t| < \e_{0}$. 
For a generic line $L_{\s}$ and generic $t \neq 0$, the intersection $\cC_{t}^{\o} \cap \Pi^{-1}(L_{\s})$ consists of an even number of points 
$\{p_{t}^{i}\}$, which converge to a finite set of points in $\cC'$ as $t \to 0$. Note that $\cC'$ separates $\cC(\e_{0})$ into 
distinct path components where $t > 0$ and $t < 0$. It seems very unlikely that such points $\{p_{t}^{i}\}$ could 
generically merge as $t \to 0$ to give a limit collection of points with different cardinality $(mod \, 2)$. We will not pursue 
this further however. 

   Note that if the conjecture \eqref{regden} that KIDs are non-generic holds, then Theorem 6.1 and \eqref{deg10} trivially also 
hold for $\e = 0$, since then $\cC_{\e_{i}} \to \cC'$ smoothly on an open-dense set. 
   
}
\end{remark}

  Next we consider the structure of the set of solutions in $\cC_{\e}$ and $\cC'$ when the restriction to the $\o$-level sets 
of $F$ is removed. Thus, we consider the map 
\be \label{wPiw}
\w \Pi_{\e}: \cC_{\e} \to \cP',
\ee
taking $(\f, X)$ to a point on the line $L_{\s} = ([g, \l \s], H)$. 
The map $\w \Pi_{\e}$ is now a smooth Fredholm map of index one. For a regular value $L_{\s}$ of $\w \Pi_{\e}$, the inverse image 
\be \label{Gamma}
\Gamma = \w \Pi_{\e}^{-1}(\{g, \s\}, H)
\ee
is a collection of curves $\Gamma = \{\ell(t)\} = \{(\f(t), X(t))\}$ mapping under $\Pi_{\e}$ to the straight line curve 
$([g, \l \s], H) \in \cD'$, $\l \in \bR$. Apriori the number of such curves (the cardinality of $\Gamma$) could be infinite. 
However, setting $\cC_{\o_{1}}^{\o_{2}} = \{(\f, X) \in \cC_{\e}: 0 < \o_{1} \leq F(\f) \leq \o_{2}\}$, the intersection 
$$\Gamma \cap \cC_{\o_{1}}^{\o_{2}},$$
consists of a finite number of compact curves, for any $\o_{1} \leq \o_{2}$. On the other hand, there are also 
data $(\{g, \s\}, H)$ for which $\Gamma = \emptyset$; this occurs for some values of $(\{g, \s\}, H)$ even when $H = const$.  

\medskip 

  The basic question is to understand when $\Gamma$ is non-empty and then further, the structure of the image 
$\Pi(\Gamma)$, for any fixed $(\{g, \s\}, H)$. We do this first in the ``small-scale" region where $\o$ is 
small and following that consider the large scale region. We also work with $\e > 0$, so that $\cC_{\e}$ is a smooth 
manifold. However, all the considerations to follow are independent of $\e$ and hold also for the $\e \to 0$ limit $\cC'$, 
cf.~the proof of Theorem 1.3 below.  

   Choose a component $\ell(t)$ of $\Gamma$ and assume without loss of generality that $t \in \bR$. Its image under $\Pi$ 
then determines $\l$ as a function of the parameter $t$, so that 
\be \label{lt}
\l = \l(t).
\ee
The properness from Theorem 5.1 implies that $\o(t) = \o(\ell(t))$ covers the full range 
\be \label{full}
\o(t) \in (0, \infty).
\ee
For if $\o(t) \to \o_{0}$ as $t \to \pm \infty$. then the limit of the curve $\ell(t)$ exists in $\cC_{\e}$ and so can be continued, 
contradicting the fact that $\ell(t)$ is maximal, i.e.~a component of the full inverse image of $L_{\s}$.

\medskip 

  In the following results, we show that the small $\o$, $\l$ behavior discussed in the CMC case in \S 5 is stable under large scale 
deformation into the far-from-CMC regime when $Y[g] \neq 0$. 

\begin{theorem} 
If $Y[g] > 0$, then for any given $L_{\s} = ([g, \l \s], H) \in \cD'$, there are constants $\o_{0} > 0$ and $\l_{0} > 0$, 
such that for any $\o$ satisfying 
$$0 < \o \leq \o_{0},$$
there is a solution $(\f, X) \in \cC_{\e}^{\o}$ with $\Pi(\f, X) = ([g, \l \s], H)$, for any $\e > 0$. Moreover, for such solutions, 
$$0 < |\l| < \l_{0}.$$

  If $Y[g] < 0$, then for any given $L_{\s} = ([g, \l \s], H) \in \cD'$, there is a constant $\o_{0} > 0$, such that 
there are no solutions $(\f, X) \in \cC_{\e}^{\o}$ with $\Pi(\f, X) = ([g, \l \s], H)$, for any $\l$, for any $\o \leq \o_{0}$ 
and for any $\e$ sufficiently small. 
\end{theorem}

{\bf Proof:} The second statement is just a restatement of Proposition 5.5, so we assume $Y[g] > 0$. 
Given a point  $(\{g, \s\}, H) \in \cP'$ with $Y[g] > 0$, choose a constant $H_{0}$ and consider a path 
$z(s)$ in $\cP'$, $s \in [0,1]$ from  $(\{g, \s\}, H_{0})$ to $(\{g, \s\}, H)$, e.g. 
\be \label{zt}
z(s) = (\{g, \s\}, (1-s)H_{0} + s H).
\ee
Choose a value $\o$ (small) and consider the smooth proper Fredholm map $\Pi_{\e}^{\o}: \cC_{\e}^{\o} \to \cD'$. 
The path $z(s)$ may be perturbed slightly if necessary, keeping endpoints fixed, so that $z(s)$ is transverse to $\Pi_{\e}^{\o}$. 
The inverse image $(\Pi_{\e}^{\o})^{-1}(z(t))$ is then a finite collection of curves, i.e.~either circles $S^{1}$ or arcs $\sim I$ with endpoints in 
the fiber $(\Pi_{\e}^{\o})^{-1}(z(0) \cup z(1))$ over the endpoints $z(0)\cup z(1)$. By the existence and uniqueness for CMC solutions 
as discussed in \S 2, the fiber $(\Pi_{\e}^{\o})^{-1}(z(0))$ over $z(0)$ consists of two points $(g, K_{\pm})$ with solution $(\f, 0)$ over 
the pair $([g, \pm \l \s], H_{0}) \in \cD'$. 

  There are now two possibilities. Namely, either the two points $(g, K_{\pm})$ over $z(0)$ are connected by an arc $I \subset 
(\Pi_{\e}^{\o})^{-1}(z(s))$, so (say) $I(0) = (g, K_{-})$, $I(1) = (g, K_{+})$, or if not, then there are a pair of arcs $I_{\pm} \subset 
(\Pi_{\e}^{\o})^{-1}(z(s))$ with $I_{\pm}(1) \in (\Pi^{\o})^{-1}(z(1))$. In the former case, since $\l(0) = -\l < 0$ and $\l(1) = \l > 0$, 
there must exist $s_{0} \in I$ such that $z(s_{0})$ maps to $([g, 0], (1-s_{0}H_{0} + s_{0}H)$ under $\Pi_{\e}^{\o}$. However, 
Proposition 5.4 shows this is not possible; $\l$ is bounded away from zero on $\cC_{\e}^{\o}$ for $\o$ sufficiently small, 
given such control on $([g, \s], H)$.  

  It follows that the second case above holds, which gives the existence of two distinct solutions $(\f_{\pm}, X_{\pm}) \in \cC_{\e}^{\o}$ over 
$(\{g, \s\}, H)$ with $|\l| > 0$ small and $\l$ of opposite signs. 

{\endproof} 

  Theorem 6.3 does not rule out the possibility of more than two solution curves over fixed data in $\cD'$ when $Y[g] > 0$. 
This is addressed in the next result. 

\begin{proposition}
For $Y[g] > 0$, $\o_{0}$ as in Theorem 6.3 and for $\l(t)$ as in \eqref{lt} one has 
$$\l'(t) \neq 0,$$
when $\o \leq \o_{0}$. Further, for a given $\l$ with $|\l| < \l_{0}$ there is a unique solution $(\f, X)$ in $\cC_{\e}^{\o}$, 
$\o \leq \o_{0}$ with data $([g, \l \s], H)$. 
\end{proposition} 

{\bf Proof:} Let $(\f', X')$ be the $t$-derivative of a curve $\ell(t) \in \Gamma \cap \cC_{0}^{\o_{0}}$. Since $g_{0}$, $\s$ and $H$ 
are fixed, the linearization of the divergence constraint \eqref{div} gives 
$$\d ({\tfrac{1}{2N}}\d_{0}^{*}X') = -4\f^{5}\f' dH,$$
while the linearization of the Lichnerowicz equation gives 
\be \label{long}
8\D \f' = R_{0}\f' + {\tfrac{2}{3}}H^{2}\f^{4}\f' + 7|\l \s + {\tfrac{1}{2N}}\d_{0}^{*}X|^{2}\f^{-8}\f'
\ee
$$ - \f^{-7}[\l \l' |\s|^{2} + ({\tfrac{1}{2N}})^{2}\<\d_{0}^{*}X', \d_{0}^{*}X\> + \l'{\tfrac{1}{2N}}\<\s, \d_{0}^{*}X\> + \l
{\tfrac{1}{2N}}\<\s, \d_{0}^{*}X'\>].$$
(Here we have dropped the terms $(\mu, \xi)$ from the expression in \eqref{long}; they enter in a way which makes no 
difference in the argument to follow). Pair \eqref{long} with $\f'$ and integrate over $M$. Then the left side of \eqref{long} is 
negative, while all terms except the last bracketed term on right are positive. For $\o$ small and $Y[g] \geq 0$, by \eqref{ll0} one 
has $\l \sim O(\f^{4})$, $\d_{0}^{*}X = O(\f^{6})$ and $\d_{0}^{*}X' = O(\f^{5})O(\f')$. Suppose then $\l' = 0$. The term 
$\<\d_{0}^{*}X', \d_{0}^{*}X\>$ is then $O(\f^{4})O(\f')\f'$ while last term $\l\<\s, \d_{0}^{*}X'\>$ is then $O(\f^{2})O(\f')\f'$. 
Both of these are small compared with the $R_{0}(\f')^{2}$ term, and so one must have $\f' = 0$. It follows then also that $X' = 0$. 

   The proof of uniqueness is the same, using the difference $\f_{1}-\f_{2}$ and $X_{1} - X_{2}$ 
in place of the (limit of) the difference quotient. 

{\endproof}

   In the following, we choose the direction of parameter $t$ so that $\l'(t) > 0$ for $t \sim -\infty$, so for $\l$ small.  
The next result summarizes the work above, proving also that $\o$ increases monotonically with $\l$ for $\l$ small. 
This is the $\cC_{\e}$ version of Theorem 1.3, for regular values $L_{\s}$. 

\begin{corollary}
For $Y[g] > 0$, and for each $\s$ with $|\s|_{C^{m-1,\a}} = 1$, there is $\l_{0} > 0$, depending only on $([g], H)$, such that 
there is a unique solution $(\f, X)$ of the constraint equations with $\Pi(\f, X) = ([g, \l \s], H)$ with $\o$ small, for each 
$$\l \in (0, \l_{0}].$$
Further, $\l$ is a smooth, monotonically increasing function of $\o$ for $\o \leq \o_{0}$. 
\end{corollary}

{\bf Proof:} It suffices to prove the last statement. For this, evaluate \eqref{long} at a point $p$ realizing $\min \f'$. 
The left side of \eqref{long} is then non-negative, while $\l' > 0$ also. A brief inspection shows that if $\min \f' < 0$, 
then all the main terms on the right are negative. In fact, the only term which is not obviously negative (or of lower order) 
is the term $\f^{-7}\l'\frac{1}{2N}\<s, \d_{0}^{*}X\>$. But $\d_{0}^{*}X = O(\f^{6}) < < \l$ by \eqref{ll0}, so this term is 
small compared with the $\f^{-7}\l \l'|\s|^{2}$ term. 

  It follows that $\min \f' > 0$, which easily implies $\o' > 0$. Thus $\o(t)$ is strictly increasing with $t$ for 
$t \sim -\infty$ and so also strictly increasing with respect to $\l$. 

{\endproof}

\noindent 
{\bf Proof of Theorem 1.3.} 

   It is now straightforward to complete the proof of Theorem 1.3. Thus Corollary 6.5 proves Theorem 1.3 for generic $\e > 0$ 
and generic lines $L_{\s}$. As noted above, the work in \S 5 and \S 6, in particular the compactness properties related to Theorem 
5.1, (cf.~also Remark 5.7), implies that the behavior described in Theorem 6.3 - Corollary 6.5 is stable when passing to limits $\e \to 0$ 
and to arbitrary $L_{\s} \in \cP'$. (Here we recall that regular values $L_{\s}$ are open and dense in $\cP'$, by the Sard-Smale 
theorem). This proves Theorem 1.3 in general. 

{\endproof}

\begin{remark}
{\rm Theorem 6.3 shows that Corollary 6.5 does not hold in the region $Y[g] < 0$; there is a transition from 
existence to non-existence as $Y[g] > 0$ passes through $Y[g] = 0$ to $Y[g] < 0$ when $\o$ is sufficiently small. 
One sees this most clearly when passing through the region $\cC^{cmc}$. 

  Thus, recall from Theorem 2.1 that all points in $\cC^{cmc}$ are regular points and $\Pi$ is a diffeomorphism 
in a neighborhood of $\cC^{cmc}$. Consider a pair of paths $z_{\pm}(s) = ([g_{s}, \pm \l \s], H_{s}) \in \cD'$, $\l \neq 0$, 
with conformal classes $[g_{s}]$ satisfying 
$$Y[g_{0}] > 0, \ Y[g_{1}] < 0, \ {\rm with} \ Y[g_{\frac{1}{2}}] = 0 \  {\rm and} \  H_{\frac{1}{2}} = const.$$ 
Assume also $\l$ is sufficiently small. Then for $s \in [0, \frac{1}{2}]$ the paths $z_{\pm}(s)$ lift uniquely to a pair of 
curves $(\f(s), X(s)) \in \cC_{0}^{\o_{0}}$, $\o_{0}$ small, solving the constraint equations. The diffeomorphism property above 
implies that the curves $z_{\pm}(s)$ continue in $\cC_{0}^{\o_{0}}$ to some open interval containing $[0, \frac{1}{2}]$. 
However, Theorem 6.3 shows that at some point $s > \frac{1}{2}$, say $s = \frac{3}{4}$, the paths $z_{\pm}(s)$ merge or 
join at $s = \frac{3}{4}$ and there is no solution $(\f, X)$ to the 
constraint equations over $z_{\pm}(s)$ in $\cC_{0}^{\o_{0}}$ for $s > \frac{3}{4}$. 

  This illustrates a clear (and generic) bifurcation or fold behavior of the map $\Pi$ in such regions. 

}
\end{remark}

  We conclude the paper with a discussion of the large-scale behavior when $\o$ becomes large, i.e.~the global behavior of 
the collection of curves $\Gamma$ in \eqref{Gamma}. 

   For any line $L_{\s} = ([g, \l \s], H_{0}) \in \cD_{+}^{cmc}$ with $H_{0} = const$, the fiber $\Gamma = 
(\w \Pi_{\e}^{-1})(\{g, \s\}, H_{0})$ consists of a pair of lines $\ell_{\pm}$ with $\l$ varying smoothly over $(-\infty, 0)$ in $\ell_{-}$ 
and $\l$ varying smoothly over $(0, \infty)$ in $\ell_{+}$. The intersections $\ell_{-}\cap \cC^{\o}$, $\ell_{+}\cap \cC^{\o}$ 
consist of an odd number of points, for generic $\o$, and are unique for $\o > > 1$ and $\o < < 1$. For $L_{\s} \in \cD_{-}^{cmc}$, 
the lines $\ell_{\pm}$ meet smoothly at a point with $\l = 0$, at a minimum value of $\o = \o_{0} > 0$. 

Given a general regular value $(\{g, \s\}, H)$ of $\w \Pi_{\e}$, as in the proof of Theorem 6.3, let $z(s)$ be a smooth path in 
$\cP'$ joining $(\{g, \s\}, H_{0})$ to $(\{g, \s\}, H)$, for example of the form \eqref{zt}, transverse to $\w \Pi_{\e}$. 
The inverse image 
$$S = \w \Pi_{\e}^{-1}(z),$$
is a properly embedded surface, not necessarily connected, in $\cC_{\e}$ with $\Gamma \subset \partial S$.  For a generic $\o$, 
the intersections $S \cap \cC_{\e}^{\o}$ are a collection of 1-manifolds, i.e.~circles $S^{1}$ or arcs $\simeq I$, with boundary 
$\partial I$ contained in the fibers $\cC_{\e}^{\o}\cap \w \Pi_{\e}^{-1}(z(0) \cup z(1))$. The $1$-manifold $S\cap \cC_{\e}^{\o}$ gives 
a cobordism between the points in the fibers $\cC_{\e}^{\o}\cap \w \Pi_{\e}^{-1}(z(0))$ and $\cC_{\e}^{\o}\cap \w \Pi_{\e}^{-1}(z(1))$. 

  It would be very interesting to study the topology of such surfaces $S$ in more detail. Their topology may well be related to 
the existence of non-trivial topology in the space $\cC_{\e}$ or $\cC'$ of solutions of the constraint equations. In this regard, 
we make only the following remarks. 

  Suppose there exists a connecting curve $z(s)$ as above and $\o \in \bR^{+}$ such that  
$$\l \neq 0$$
everywhere on $S\cap \cC_{\e}^{\o}$, so that $\w \Pi_{\e}$ is defined on $S\cap \cC_{\e}^{\o}$. It then follows just as in the proof of 
Theorem 6.2 that there exists $\l_{+}> 0$ and $\l_{-} < 0$ such that the data $([g, \l_{\pm} \s], H)$ are solvable, with 
$(\f, X) \in \cC_{\e}^{\o}$. Of course the proof of Theorem 6.3 proves this is the case for $Y[g] > 0$ and $\o$ sufficiently small, 
so that $S\cap \cC_{0}^{\o}$ consists of a pair of disjoint arcs connecting $\cC_{\e}^{\o}\cap \w \Pi_{\e}^{-1}(z(0))$ to 
$\cC_{\e}^{\o}\cap \w \Pi_{\e}^{-1}(z(1))$. 

  This shows that the existence of solutions $(\f, X)$ with $\l = 0$, i.e.~with data of the form 
$$([g], 0, H),$$
may give a possible obstruction to the existence of solutions over $(\{g, \s\}, H')$, for some $H'$ near $H$ or $[g']$ near $[g]$.  

  As noted above in \eqref{full}, on any component $\ell_{\s}$ of $\Gamma$, the function $F$ takes on all values $\o \in \bR^{+}$ 
and so is a proper function on $\ell_{\s}$. By Theorem 5.1, on $\ell_{\s}$, 
\be \label{ls}
\l \to \infty \Rightarrow \o \to \infty.
\ee
On the other hand, it is not true that $\l$ is a proper function on $\ell_{\s}$ in general, i.e.~the converse of \eqref{ls} 
may not hold. This follows from the fundamental non-existence result of Nguyen \cite{Ngu}. It is then of basic interest to 
understand the value 
$$\l_{0} = \sup \{|\l(\f, X)|: \w \Pi(\f, X) = (\{g, s\}, H)\}.$$

This leads to the ``limit equation" introduced by Dahl-Gicquaud-Humbert \cite{DGH}. We give a simple derivation here 
for completeness, cf.~also \cite{GN}. 
\begin{proposition}
Let $K$ be a compact set in $\cP'$ with $H > 0$ and let $(\f_{i}, X_{i})$ be a sequence of solutions 
of the constraint equations with $\w \Pi(\f_{i}, X_{i}) \in K$. If $F(\f_{i}) \to \infty$ but 
$$\sup_{i}|\l(\f_{i}, X_{i})| < \infty,$$
then there is a $C^{2,\a}$ solution $\bar X$ to the limit equation 
\be \label{limit}
\d ({\tfrac{1}{2N}}\d_{0}^{*}\bar X) = -\sqrt{{\tfrac{2}{3}}}|{\tfrac{1}{2N}}\d_{0}^{*}\bar X|\frac{dH}{H}.
\ee
At points where $\d_{0}^{*}\bar X \neq 0$, the solution $\bar X$ is $C^{m,\a}$. 
\end{proposition}

{\bf Proof:} Let $m_{i} = \sup \f_{i}$, so that $m \to \infty$. As in \eqref{renorm}, renormalize the divergence constraint \eqref{div} 
by dividing by $m^{6}$ to obtain
\be \label{limit1}
\d ({\tfrac{1}{2N}}\d_{0}^{*}\bar X) = \bar \f^{6} dH,
\ee
where $\bar X = X/m^{6}$, $\bar \f = \f / m$ and we have dropped the subscript $i$ for simplicity. It follows from 
elliptic regularity that $\bar X$ remains uniformly bounded in $C^{1,\a}$ and so $\d_{0}^{*}X$ is uniformly bounded 
in $C^{\a}$. Similarly, renormalizing the Lichnerowicz equation \eqref{lich} gives 
$$8m^{-4}\D \bar \f = m^{-4}R_{0}\bar \f - |\bar \s + {\tfrac{1}{2N}}\d_{0}^{*}\bar X|^{2}\bar \f^{-7} + {\tfrac{2}{3}}H^{2}\bar \f^{5},$$
where $\bar \s = \s / m^{6} \to 0$. Since $m^{-4}\D \bar \f \to 0$ weakly (i.e.~as a distribution) and $m^{-4}R_{0}\bar \f \to 0$ 
in $L^{\infty}$, it follows that, after passing to a subsequence, that there is a limit $(\bar \f, \bar X)$ satisfying  
\be \label{limit2}
|{\tfrac{1}{2N}}\d_{0}^{*}\bar X|^{2}\bar \f^{-7} = {\tfrac{2}{3}}H^{2}\bar \f^{5},
\ee
weakly, i.e.~as distributions. Since the right side of \eqref{limit} is in $L^{\infty}$, the left side is also and hence, multiplying 
by the $L^{\infty}$ function $\bar \f$ gives  
$$|\d_{0}^{*}\bar X|^{2} = {\tfrac{2}{3}}H^{2}\bar \f^{12},$$
in $L^{\infty}$. Substituting this in \eqref{limit1} gives \eqref{limit}. Bootstrapping via elliptic regularity in the usual way gives 
$\bar X \in C^{m,\a}$, where $|\d_{0}^{*}\bar X| \neq 0$. Near points where $|\d_{0}^{*}\bar X| = 0$, the function 
$|\d_{0}^{*}\bar X|$ is Lipschitz, so in $C^{\a}$, so that $\bar X \in C^{2,\a}$. 

{\endproof} 

   This leads naturally to the following result. 
\begin{corollary}
Let $\O$ be a domain in $\cG'\times c^{m,\a}(M)$ with $Y[g] > 0$ and $H > 0$ and suppose the 
limit equation \eqref{limit} has no non-zero solution for $([g], H) \in \O$. 

  Then for any $\s \neq 0$ there is a solution $(\f, X)$ of the constraint equations \eqref{div}-\eqref{lich} over the data 
$([g, \s], H)$ with $([g], H) \in \O$. Further, generically the number of solutions is finite and odd. 
\end{corollary}

{\bf Proof:} Let $\cD_{\O}' \subset \cD'$ be the bundle region over $\O$. Proposition 6.7 implies that whenever $F \to \infty$ 
in the region $\cC_{\O}' = (\Pi')^{-1}(\cD_{\O}')$ then necessarily $\l \to \infty$ also. This implies that 
\be \label{prop1}
\Pi': \cC_{\O}' \to \cD_{\O}',
\ee
is a continuous proper map; compare with Theorem 5.1. Similarly, the $\e$-perturbation 
\be \label{prop2}
\Pi_{\e}': \cC_{\O, \e}' \to \cD_{\O}',
\ee
is a smooth proper Fredholm map, of Fredholm index $0$. Hence the map \eqref{prop2} has a well-defined Smale degree. We claim 
that, for any $\e$ sufficiently small,  
\be \label{deg1}
deg_{\bZ_{2}}\Pi_{\e}' = 1.
\ee
To see this, choose a regular value $([g, \l \s], H)$, $|\s|_{C^{m-1,\a}} = 1$, of $\Pi_{\e}'$. By Corollary 6.5, for $\l$ 
sufficiently small, there is a unique solution $(\f, X)$ of the $\e$-perturbed constraint equations with $\Pi_{\e}'(\f, X) = 
([g, \l \s], H)$. Since $(\f, X)$ is the unique regular point over the regular value $([g], \l \s, H)$, this proves \eqref{deg1}. 

  Using the compactness results from \S 5 and \S 6 as in the proof of Theorem 1.3, one may pass to the limit $\e \to 0$ to 
obtain a solution $(\f, X)$ of the vacuum constraint equations \eqref{div}-\eqref{lich} over $([g, \s], H)$, for any $\s \neq 0$. 
The last statement then also follows from \eqref{deg1}.  

{\endproof}

  Note that the basic non-existence result of Nguyen \cite{Ngu} implies that $\O$ is not all of $\cG'\times c^{m,\a}(M)$ and 
so it remains an interesting open problem to characterize such domains $\O$.  

\begin{remark} 
{\rm As is well-known, we note that the scaling transformation \eqref{scale} transforms solutions $(\f, X)$ with small value 
$\o$ of $F$ to those with large $\o$, while also increasing the values of $\s$, $X$ but decreasing the mean curvature $H$. 
Thus for example, as discussed in \cite{GN}, the existence and uniqueness result given in Theorem 6.3 corresponds to a similar result 
for $H$ close to $0$. 

}
\end{remark}

\bibliographystyle{plain}

\end{document}